\documentclass[letterpaper,pra,twocolumn]{revtex4}
\usepackage[latin1]{inputenc}
\usepackage{amsmath,amsbsy}
\usepackage{amsfonts,hyperref}
\usepackage{bbm}
\usepackage{verbatim}
\usepackage{amsthm}

\newcommand{\ket}[1]{|#1\rangle}
\newcommand{\bra}[1]{\langle #1|}
\newcommand{\bracket}[2]{\langle #1|#2\rangle}
\newcommand{\ketbra}[1]{|#1\rangle\langle #1|}
\newcommand{\cket}[1]{|\widetilde{#1}\rangle}

\newcommand{\eps}{\epsilon}
\newtheorem{theorem}{Theorem}
\newtheorem{lemma}{Lemma}
\newtheorem{corollary}{Corollary}

\newcommand{\jmr}[1]{#1}
\newcommand{\m}[1]{\mathbf{#1}}
\newcommand{\ba}{{\boldsymbol{\alpha}}}
\newcommand{\bb}{{\boldsymbol{\beta}}}
\newcommand{\bl}{{\boldsymbol{\lambda}}}
\newcommand{\bm}{{\boldsymbol{\mu}}}
\usepackage{amsmath}

\def\Tr{{\mathrm{Tr}}}

\newcommand{\leqslant}{\leq}

\begin{document}
\author{Joseph M.~Renes$^1$ and Jean-Christian Boileau$^{2}$}
\affiliation{$^1$Institut f\"ur Angewandte Physik, Technische Universit\"at Darmstadt, Hochschulstra\ss e~4a, 64289 Darmstadt, Germany\\
$^2$Center for Quantum Information and Quantum Control, University of Toronto, Toronto, ON, Canada M5S 1A7}

\title{Physical underpinnings of privacy}

\begin{abstract}
One of the remarkable features of quantum mechanics is the ability to ensure secrecy.
Private states embody this effect, as they are precisely those multipartite quantum
states from which two parties can produce a shared secret that cannot under any 
circumstances be correlated to an external system. Naturally, these 
play an important role in quantum key distribution (QKD) and quantum information theory. However, a general distillation method has heretofore been missing. Inspired by Koashi's complementary control scenario [M.~Koashi, e-print arXiv:0704.3661 (2007)], we give a new definition of private states in terms of one party's potential knowledge of two complementary measurements made on the other and use this to construct a general method of private state distillation using quantum error-correcting codes. The procedure achieves the same key rate as recent, more information-theoretic approaches while demonstrating the physical principles underlying privacy of the key. Additionally, the same approach can be used to establish the hashing inequality for entanglement distillation, as well as the direct quantum coding theorem.
\end{abstract}
\maketitle

\section{Introduction}
Appeal to physical concepts such as the uncertainty principle and entanglement formed the basis of the original security proofs of quantum key distribution (QKD). An uncertainty relation between complementary observables inspired the first, Mayers's security proof of the BB84 protocol~\cite{mayers_quantum_1996}. Later, building on arguments from Lo and Chau~\cite{lo_unconditional_1999}, Shor and Preskill~\cite{shor_simple_2000} showed how BB84 could be understood as a virtual entanglement distillation protocol, thereby using the monogamy of entanglement to ensure the privacy of the key. This method subsequently found wide application not only to specific~\cite{lo_proof_2001, tamaki_unconditionally_2003, boileau_unconditional_2005, tamaki_unconditionally_2006} and generic~\cite{renes_generalized_2006} ideal protocols, but also to protocols including a description of realistic devices~\cite{gottesman_security_2004}. 
Recently, Koashi combined the two \jmr{methods}~\cite{koashi_unconditional_2006} and formulated a simple security proof for BB84 with uncharacterized detectors~\cite{koashi_efficient_2006}. 

A somewhat different, more information-theoretic approach adapts classical schemes of extracting secret bits from partially private data to the case in which the eavesdropper holds quantum information. 
If $X$, $Y$, and $Z$ are classical random variables held by two honest parties Alice and Bob, along
with an eavesdropping third party, Eve, then a result by Csisz\'ar and K\"orner states that by one way communication from Alice to Bob the honest parties can extract a key at a rate of $I(X{:}Y)-I(X{:}Z)$ bits from asymptotically many such random variables~\cite{csiszar_broadcast_1978}. Devetak and Winter showed how to distill secret keys from tripartite quantum states at the quantum version of this rate, obtained by replacing Bob's and Eve's classical random variables with quantum states~\cite{devetak_distillation_2005}. Building on a result by Renner and K\"onig~\cite{renner_universally_2005}, Kraus, Gisin, and Renner established the security of generic QKD protocols \jmr{operating at this rate} using arbitrary universal hash functions~\cite{kraus_lower_2005, renner_information-theoretic_2005, renner_security_2006}. 

The essential difference between the two approaches lies in the basis of privacy and the treatment of the eavesdropper. In the latter, privacy is established  directly. Alice and Bob employ privacy amplification to eliminate any information Eve may have about their prospective \emph{classical} key, even if she holds quantum information. This general approach works in any kind of cryptographic setting, classical, quantum, or otherwise, provided Alice and Bob have some estimate of Eve's information. In the quantum setting, this estimate can be obtained by assuming Eve holds the purification of the quantum state held by Alice and Bob; that this limits her information is the reason QKD is possible from this point of view. 

\jmr{In the former approach, the honest parties} no longer concern themselves with the details of the eavesdropper, but instead concentrate on creating a \emph{quantum} state that can produce a secret key when appropriately measured. For example, maximal entanglement will ensure privacy of a key generated in any basis by the monogamy property mentioned above. Entanglement is sufficient \jmr{for this purpose, but unnecessary; the broader} class of states suitable for creating keys are termed \emph{private states}~\cite{horodecki_secure_2005}. These are closely related to maximally entangled states, but may also include additional systems, collectively called the \emph{shield}. The shield does not contribute directly to the key, but, as the name suggests, serves to block its correlations from would-be eavesdroppers. From this perspective, the success of QKD hinges \jmr{on the existence of quantum correlations which} implies that the results of certain measurements are completely secret.

Each approach has its advantages. The physical picture is perhaps more intuitive, tracing the origins of privacy to physical concepts such as entanglement, complementarity, and the uncertainty principle. On the other hand, the information-theoretic approach has led to more general proofs with higher lower bounds and lower upper bounds on the secret key rate~\cite{devetak_distillation_2005, kraus_lower_2005, renner_information-theoretic_2005, renner_security_2006}. 

These results, specifically rates of secret key distillation, have also been used to derive some of the central results of quantum information theory, namely the hashing inequality on the asymptotic rate of entanglement distillation and the direct quantum coding theorem for the quantum channel capacity. In principle, it should be possible to arrive at the same results in the physical picture, as every key distillation protocol in principle leads to a private state distillation protocol by performing the operations coherently~\cite{horodecki_general_2005}. Put differently, the results from the information-theoretic viewpoint can be used to construct such distillation protocols, but these have not yet been fully understood from the more physical point of view. 

We provide the missing piece of the puzzle in this paper by formulating a new characterization of private states based on the uncertainty principle and using this to construct a protocol using Calderbank-Shor-Steane (CSS) codes~\cite{calderbank_good_1996, steane_multiple-particle_1996}, which distills private states at the quantum Csisz\'ar-K\"orner rate. The essential idea is that if and only if measurements on Alice's key system in either one of two conjugate bases can be perfectly predicted by the other systems available to the honest parties, then the joint state is a private state and Eve can have no correlation with the key. In particular, Bob's key system should be perfectly correlated with Alice's, while the shield may be used to predict her conjugate observable. 

Here, privacy of the key rests on quantum-mechanical complementarity, since the fact that either of the conjugate observables could be predicted by the honest parties means that Eve has no correlation with either. This echoes the recent result by Koashi showing that secret key distillation is equivalent to a protocol involving complementary measurements he termed complementary control~\cite{koashi_complementarity_2007}, and indeed our work is inspired by these results. 

By explicitly including Bob and the shield into the analysis, the means of private state distillation become clear: Alice merely needs to reveal some information about her key system such that the other systems could in principle predict both measurements. We shall demonstrate how the syndromes of a CSS code are ideally suited for this purpose, and that the resulting distillation protocol essentially amounts to applying a slightly modified Holevo-Schumacher-Westmoreland (HSW) theorem~\cite{holevo_capacity_1998,schumacher_sending_1997} twice. Constructing a distillation procedure in this manner, one focused on the shared 
quantum correlations, generalizes the quantum privacy amplification method of Deutsch \emph{et al.}~\cite{deutsch_quantum_1996} and recalls 
the connection between quantum privacy and quantum coherence discovered by Schumacher and Westmoreland~\cite{schumacher_quantum_1998}.

This approach also gives a new proof of the hashing inequality, which states that the rate of one-way entanglement distillation using many copies of the state $\rho_{AB}$ is lower bounded by the coherent information $I_c(A \rangle B)=S(B)-S(AB)$ (the same lower bound applies to the extractable one-way secure key rate). As discussed in \cite{devetak_relating_2004}, this result combined with quantum teleportation provides proof of the direct quantum coding theorem, which gives a lower bound to the quantum channel capacity in terms of the coherent information. The main difference from previous proofs is that we bound Eve's information about the key by the amount of information that Bob can obtain about Alice's conjugate basis measurement, which 
then leads to an explicit construction of the decoder.

The paper is organized as follows. First we give the new characterization of private states in Sec.~\ref{exactprivatestate}, and show how quantitative statements of complementarity such as the entropic uncertainty principle of Maassen and Uffink~\cite{maassen_generalized_1988} and a related mutual information tradeoff given by Hall~\cite{hall_information_1995} imply privacy of the key. We then extend this to the case of approximate private states in Sec.~\ref{sec:ApproxPrivateState}, explaining the relation to Koashi's complementary control scenario. Section~\ref{sec:psd} presents our main results, which we divide into two parts. We first prove a one-shot distillation theorem showing how to use the structure of CSS codes for private state distillation, in a form useful as a building block for QKD security proofs. 
We then give a distillation protocol based on these ideas that achieves the quantum Csisz\'ar-K\"orner rate.  In Sec.~\ref{sec:HashIneq}, we use a coherent version of those arguments to prove the hashing inequality.  In Sec.~\ref{sec:previouswork}, we discuss relation to previous work, and we conclude in Sec.~\ref{sec:concl} with a summary and open problems.

\section{Exact Private States}
\label{exactprivatestate}

A perfect secret key shared by Alice and Bob is a uniformly distributed random variable about which the eavesdropper Eve has zero information, or more formally, $\kappa^{ABE}:=\left(\frac{1}{d}\sum_{k=0}^{d-1} P_{k}^{A}\otimes P_k^B\right)\otimes \rho^E$ for some $\rho^E$, where $P_k:=\ket{k}\bra{k}$ is the projector onto ``standard'' basis element $\ket{k}$. Note that this choice of basis is arbitrary for each system. Although we use a quantum-mechanical description, note that Alice and Bob's systems are essentially classical; states of this form are sometimes termed \emph{ccq} states to reflect this fact. 

Private states, meanwhile, are quantum states for which standard basis measurements by
Alice and Bob yield a perfect secret key. When producing a key from an alphabet of $d$ letters, the 
key registers $A$ and $B$ are $d$-dimensional quantum systems. Additionally, they 
may possess some auxiliary ``shield'' systems that are not directly involved in creating the
key. These systems are nevertheless important as they are not held by the eavesdropper and
can shield the key correlations from her. Although the shield may have several parts distributed
between Alice and Bob, here we lump them together into the system labelled $S$.

In contrast to the explicit reference to Eve's system in the definition of secret keys, the privacy of a state $\gamma^{ABS}$ can be determined solely from the systems held by Alice and Bob. The canonical example of such an effect comes from a maximally entangled state, which by virtue of the monogamy of entanglement creates secret keys upon measurement. Though there is no shield in this example, it makes the point that the quantum correlations between Alice and Bob's systems are enough to establish secrecy of the key. 

Private states are in fact closely related to maximally entangled states, as shown by~\cite{horodecki_secure_2005}. To recapitulate their result, first define a \emph{twisting operator} to be a controlled unitary of the form $U^{ABS}:=\sum_{jk} P^{A}_j\otimes P^B_k\otimes V^S_{jk}$ for any arbitrary unitaries $V^S_{jk}$. Then Theorem 1 of~\cite{horodecki_secure_2005} states that $\gamma^{ABS}$ is a private state iff it is of the form
\begin{equation}
\label{eq:twistedstated}
\gamma^{ABS}=U^{ABS}(\Phi^{AB}_d\otimes \xi^S)U^{\dagger ABS},
\end{equation}
where $\xi^S$ is an arbitrary state and $\Phi^{AB}_d$ is the density operator associated with the canonical entangled state $\ket{\Phi^{AB}_d}:=\frac{1}{\sqrt{d}}\sum_{k=0}^{d-1}\ket{kk}^{AB}$; note that actually only the $V_{kk}$ are relevant. Clearly, measurement of the $A$ and $B$ systems results in a secret key since the same key would result if the state were first untwisted, and Eve cannot distinguish the cases in which the state has been untwisted or not. 
Conversely, purifying a secret key and using the fact that Eve's marginal state is fixed along with the fact that purifications of a fixed marginal are related by unitaries on the purifying system, i.e.~Uhlmann's theorem~\cite{uhlmann_transition_1976, jozsa_fidelity_1994}, guarantees the form of Eq.~(\ref{eq:twistedstated}).  

With the help of the uncertainty principle we can formulate a different
characterization of private states that emphasizes the relation of privacy to  
complementarity and does not involve statements about Eve's system. 
Consider a hypothetical measurement by one party, say Alice, on 
her key qubit in a basis \emph{conjugate} to the standard basis.
In this context, ``conjugate'' refers to any basis whose elements give random outcomes when
measured in the standard basis. A general conjugate basis has elements 
$\cket{x}:=\frac{1}{\sqrt{d}}\sum_{k=0}^{d-1}e^{i\theta_{xk}}\ket{k}$ for some set of $\theta_{xk}\in\mathbb{R}$
such that $\frac{1}{d}\sum_{k}e^{i(\theta_{xk}-\theta_{yk})}=\delta_{xy}$.

Due to the conjugate nature of the $\ket{k}$ and $\cket{x}$ bases, complementarity 
places constraints on the predictability of both measurements. In particular, the 
entropic uncertainty relation of Maassen and Uffink~\cite{maassen_generalized_1988} states that, for an arbitrary state $\rho^A$, 
\begin{equation}
\label{eq:unc}
H(Z^A)+H(\widetilde{X}^A)\geq \log_2 d,
\end{equation}
where $Z^A$ and $\widetilde{X}^A$ are any nondegenerate observables having eigenstates $\ket{k}^A$ and $\cket{x}^A$, respectively, and $H$ is the Shannon entropy of the 
outcome probabilities, measured in bits. Hence, if the outcome of $Z$ is certain, then the measurement of $\widetilde{X}$ must be random and vice versa.

\jmr{To determine how much information is \emph{simultaneously} available, 
we can include the measurement devices
themselves in the description, following Hall and Cerf \emph{et al}.~\cite{hall_information_1995,cerf_security_2002}. Whatever information can be stored in separate devices is clearly simultaneously accessible, so consider a state $\rho^{ACD}$ and POVMs $\widetilde{\Lambda}^C$ and $\Gamma^D$ that are restricted to systems $C$ and $D$, respectively.} 
\jmr{Denoting the classical conditional entropy of $Z^A$ given the measurement result $\Gamma^D$ by $H(Z^A|\Gamma^D)$, we have}
\begin{lemma}[Complementary Information Tradeoff]
\label{lemma:cit}
For a tripartite quantum state $\rho^{ACD}$, conjugate observables ${Z^A}$ and $\widetilde{X}^A$, and arbitrary measurements  $\widetilde{\Lambda}^C$ and $\Gamma^D$, 
\begin{equation}
H({Z}^A|\Gamma^D)+H(\widetilde{X}^A|\widetilde{\Lambda}^C)\geq \log_2 d
\end{equation}
where $d={\rm dim}(A)$.
\end{lemma}
\begin{proof}
Consider arbitrary measurements $\widetilde{\Lambda}^C$  and $\Gamma^D$. Since these can be performed independently simultaneously, we can define the conditional marginal state $\rho^A_{jk}:= \Tr_{CD}[\widetilde{\Lambda}^C_j \Gamma^D_k\rho^{ACD} ]/p_{jk}$\jmr{, for $p_{jk}:= \Tr[\widetilde{\Lambda}^C_j \Gamma^D_k\rho^{ACD} ]$}. Measurements of $Z^A$ and $\widetilde{X}^A$ on each of those states must obey Eq.~(\ref{eq:unc}), which in 
the current context reads $H(Z^A|\Gamma^D{=}k,\widetilde{\Lambda}^C{=}j)+H(\widetilde{X}^A|\Gamma^D{=}k,\widetilde{\Lambda}^C{=}j)\geq \log_2 d$. Averaging over the measurement outcomes and using the fact that conditioning reduces entropy, we obtain the desired result. 
\end{proof}

Note that no restriction is placed on the ability of a single system to be correlated with two complementary Alice observables, only that the correlations not be simultaneously realized. Such is the case when $\rho^{AB}$ is maximally entangled; in the EPR state, for instance, Bob can predict either the position or momentum of Alice's system, but not both at the same time. 

The information tradeoff bears directly on the question of privacy, as conjugate information can be used to exclude the eavesdropper's information about the key. Define the key to be the outcome of Alice's observable $Z^A$, let Eve hold $D$, and suppose that system $C=BS$, i.e.~the remainder of the systems under Alice and Bob's control. Then if some measurement $\widetilde{\Lambda}^{BS}$ of the $BS$ subsystem can predict the outcome of Alice's conjugate basis observable $\widetilde{X}^A$, Eve can have no information about the key: $H(\widetilde{X}^A|\widetilde{\Lambda}^{BS})=0$ implies $H(Z^A|\Gamma^E)=\log_2 d$.
Thus, complementarity assures privacy of the secret key without directly making statements about Eve's system. This line of thought leads to the new characterization of private states:
\begin{theorem}[Exact Private States]
\label{th:exactps}
$\gamma^{ABS}$ is a private state with (nondegenerate) key observables $Z^A$ and $Z^B$ iff for some 
measurement $\widetilde{\Lambda}^{BS}$
\begin{align}
\text{\emph{(a)}}&\quad H(Z^A|Z^B)=0,\quad\text{and}\\
\text{\emph{(b)}}&\quad H(\widetilde{X}^A|\widetilde{\Lambda}^{BS})=0.
\end{align}
\end{theorem}
\begin{proof}
Start with the reverse (if) implication and suppose $\gamma^{ABS}$ 
satisfies the two conditions. By the above argument, condition (b) implies $H(Z^A|\Gamma^E)=\log_2 d$ and therefore $H(Z^A)=\log_2 d$, 
whence Eve's marginal states must be independent of the key. As (a) implies the key is perfectly correlated, 
$\gamma^{ABS}$ must be a private state.

To prove the forward (only if) implication, we construct the measurement $\widetilde{\Lambda}^{BS}$ from the twisting operator $U^{BS}=\sum_k P^B_k\otimes V^S_{kk}$.
First, condition (a) follows immediately for $\gamma^{ABS}$ a private state. 
The joint probability for the conjugate measurement is given by  
\begin{align*}
p_{xy} &={\rm Tr}[\gamma^{ABS}\widetilde{P}_x^A\otimes \widetilde{\Lambda}_y^{BS}]\\
&=\frac{1}{d^2}\sum_{jk}e^{i(\theta_{xk}-\theta_{xj})}{\rm Tr}\left[\left(\ket{j}\bra{k}^B\otimes V_{jj}^S\xi^S V_{kk}^{\dagger S}\right)\widetilde{\Lambda}_y^{BS}\right]\\
&=\frac{1}{d^2}\sum_{jk}e^{i(\theta_{xk}-\theta_{xj})}{\rm Tr}\left[\left(\ket{j}\bra{k}^B\otimes \xi^S\right)U^{\dagger BS}\widetilde{\Lambda}_y^{BS}U^{BS}\right]\\
&=\frac{1}{d}{\rm Tr}\left[\left(\widetilde{P}^{*B}_x\otimes \xi^S\right)U^{\dagger BS}\widetilde{\Lambda}_y^{BS}U^{BS}\right],
\end{align*}
where $\widetilde{P}_y^{*B}$ is the conjugate of $\widetilde{P}_y^{B}$ in the standard basis. 
Condition (b) follows by setting $\widetilde{\Lambda}^{BS}_y:=U^{BS}\left(\widetilde{P}_y^{*B}\otimes \mathbbm{1}^S\right)U^{\dagger BS}$ so that $p_{xy}\propto \delta_{xy}$.
\end{proof}
From this viewpoint, privacy of the key follows from the ability of one part of the honest players' systems to predict either the key or a complementary observable of the other part; here we focused on Alice's system, but clearly the same result holds for Bob's. 

\section{Approximate Private States}
\label{sec:ApproxPrivateState}
Of course, a realistic QKD protocol can never produce a perfect secret key or a perfect private state and instead strives to create a good approximation. But what is a good approximation? Because the key is meant to be used in arbitrary further cryptographic applications, the definition of approximate must be \emph{composable} so that security statements about a whole cryptographic process can be made by individually examining the constituent parts. In this framework, a sufficient notion of approximate 
secrecy is furnished by the probability that the actual key could be distinguished from an exact secret key.
According to Helstrom's theorem~\cite{helstrom_quantum_1976}, the probability of distinguishing between the two quantum states $\rho$ and $\sigma$ is bounded by $\frac{1}{2}+\frac{1}{4}\Tr\big|\rho-\sigma\big|$. Hence the trace distance $\frac{1}{2}\Tr\big|\rho-\sigma\big|$ is the important quantity. This motivates the definition that a shared $\eps$-secret key, where $\eps$ is called the \emph{security parameter}, is any $\rho^{ABE}$ that satisfies $\Tr |\rho^{ABE}-\kappa^{ABE }| \leq 2 \epsilon$ for some perfect secret key $\kappa^{ABE}$~\cite{renner_universally_2005, ben-or_universal_2005}.

We could analogously define $\eps$-private states to be states that are $\eps$-close to exact private states in trace distance. These will lead to $\eps$-secret keys since the measurement that creates the key is a quantum operation, and the trace distance can only decrease under quantum operations. However, the converse is not true: States not $\epsilon$-close to a private state may nevertheless still generate $\eps$-secret keys. Hence a better approach is simply to say that $\psi^{ABS}$ is an $\eps$-private state when the key measurement leads to an $\eps$-secret key, with the eavesdropper system $E$ defined as any purifying system of $\psi^{ABS}$.

Intuitively, the new characterization of exact private states should be extendible to the approximate case; if Alice's key and conjugate measurements are almost perfectly predictable by the $BS$ systems, then the shared state ought to produce a good approximation of a secret key. Defining ``almost perfect predictability'' in terms of nearly zero conditional entropy, or equivalently nearly maximal mutual information, will not suffice, as this approach is not composable~\cite{gottesman_proof_2003}. Instead, the following two theorems show that an alternate definition of approximate private states can be given in terms of concrete measurements having small probabilities of error. The first 
says that if Bob is able to distinguish Alice's state measured in either one of two conjugated bases, then they share an $\eps$-private state, while the second is the converse. Only the first theorem is needed when constructing a security proof, but we provide both for completeness and to highlight the connection between our framework and Koashi's complementarity control scenario~\cite{koashi_complementarity_2007}.

\begin{theorem}
\label{TheoremMeasImplySecKey}
A state $\psi^{ABS}$ with nondegenerate key observables $Z^A$ and $Z^B$ is an $(\eps_z+\sqrt{\eps_x})$-private state if there exists
a conjugate observable $\widetilde{X}^A$ and corresponding measurement $\widetilde{\Lambda}^{BS}$ such that
\begin{align}
	p_{\rm e}&=\sum_{j\neq k}\Tr\left[(P^A_j\otimes P^B_k)\psi^{ABS}\right]\leq \eps_z,\\
\widetilde{p}_{\rm e}&=\sum_{x\neq y}\Tr\left[(\widetilde{P}^A_x\otimes \widetilde{\Lambda}_y^{BS})\psi^{ABS}\right]\leq \eps_x.
\end{align} 
\end{theorem}

\begin{theorem}
\label{thm:goodkcmexist}
If $\psi^{ABS}$ is an $\epsilon$-private state with nondegenerate key observables $Z^A$ and $Z^B$, 
then for any conjugate observable $\widetilde{X}^A$ there exists a corresponding measurement $\widetilde{\Lambda}^{BS}$ such that 
\begin{align}
	p_{\rm e}&=\sum_{j\neq k}\Tr\left[(P^A_j\otimes P^B_k)\psi^{ABS}\right]\leq \eps,\\
\widetilde{p}_{\rm e}&=\sum_{x\neq y}\Tr\left[(\widetilde{P}^A_x\otimes \widetilde{\Lambda}_y^{BS})\psi^{ABS}\right]\leq 2\eps-\eps^2.
\end{align}
\end{theorem}
\noindent As the proofs are somewhat technical, we defer them to Appendix~\ref{app:proofs}.

\section{Private State Distillation}
\label{sec:psd}

With this characterization of approximate private states, it becomes simple to construct a procedure to distill private states from an arbitrary input. Alice simply needs to reveal enough information about her system so that the states of the $B$ and $BS$ systems can be reliably distinguished. The amount of information she must reveal depends on the details of the state, and no useful answer can be given in the general case. But when Alice and Bob share asymptotically many copies of an arbitrary state ${\psi}^{ABS}$, two applications of the HSW theorem give the distillation rate, which we show equals the quantum Csisz\'ar-K\"orner rate. 

However, this distillation scenario contains the additional subtlety that the information Alice needs to reveal ostensibly comes from noncommuting measurements. Avoiding this problem is where CSS error-correcting codes come into play, as they enable the side information
to be properly defined in terms of commuting variables and also define the form of the key system of the distilled state. CSS codes were used by Shor and Preskill~\cite{shor_simple_2000} in their proof of the BB84 protocol for precisely
the same purpose, and the following distillation scheme can be understood as an extension of this method to arbitrary private states. 
This section contains the \jmr{} main results of this paper, which for clarity are subdivided into two parts: How the CSS codes enable distillation when Alice's state has dimension $d^n$, and at what rate can private states be distilled from many copies of an arbitrary resource state.

\subsection{One-shot distillation}

First we recall a few facts about CSS codes. A CSS code encoding $n-m_z-m_x$ qudits into $n$ is defined by a set of $m_z+m_x$ (commuting) stabilizer operators, 
$m_z$ operators of the form $Z^{\bf s}=Z^{s_1}\otimes Z^{s_2} \otimes \cdots \otimes Z^{s_n}$ for $0 \leq s_i \leq d-1$, and $m_x$ of the form $X^{\bf t}=X^{t_1}\otimes X^{t_2} \otimes \cdots \otimes X^{t_n}$ for  $0 \leq t_i \leq d-1$. We have implicitly 
used the definition ${\bf s}=(s_1,\dots,s_n)$ and the notation that an operator raised to a string is simply the product of the operators raised to the elements of the string. To simplify notation, we adopt the following:  
$\ket{\m{k}}=\ket{k_1}\otimes\cdots\otimes \ket{k_n}$, $\ket{\varphi_\m{k}}=\ket{\varphi_{k_1}}\otimes\cdots\otimes \ket{\varphi_{k_n}}$, and $P_\m{k}$ for $P_{k_1}\otimes \cdots\otimes P_{k_n}$ and similarly for $\widetilde{P}_\m{x}$ in the conjugate basis. 

The first operator set, the $Z$-type stabilizers, defines a code correcting errors in the standard basis (dit errors, or amplitude errors), while the second, the $X$-type stabilizers, defines a code correcting phase errors. Here, and henceforth, the operators $X$ and $Z$ are the generalized Pauli operators in $d$ dimensions~\cite{gottesman_encoding_2001}, 
given by $Z:=\sum_{k=0}^{d-1}\omega^k \ketbra{k}$ and $X:=\sum_{k=0}^{d-1}\ket{k{+}1}\bra{k}= \sum_{k=0}^{d-1}\omega^{-k} \ketbra{\widetilde{x}}$, where $\omega := e^{\frac{2\pi i}{d}}$. 

Measuring the stabilizers yields the amplitude and phase syndromes $\boldsymbol{\alpha}=(\alpha_1,\dots,\alpha_{m_z})$ and $\boldsymbol{\beta}=(\beta_1,\dots,\beta_{m_x})$, to which we associate projectors 
$\Pi_\ba$ and $\widetilde{\Pi}_\bb$, respectively. 
Since the stabilizers are products of $Z$s or $X$s, these projectors can be expressed as $\Pi_\ba=\sum_{\m{k}\in [\ba]}P_\m{k}$ 
and $\widetilde{\Pi}_\bb=\sum_{\m{x}\in [\bb]}\widetilde{P}_\m{x}$.
Meanwhile, the $[\ba]$ and $[\bb]$ are equivalence classes of standard and conjugate basis states that all share the syndromes $\ba$ and $\bb$, respectively. 

Commuting with the stabilizers (but not included in them) are the logical or encoded operators $\bar{Z}_j$ and $\bar{X}_j$, one pair for each of the $n-m_z-m_x$ encoded qudits. Crucially, these may also be chosen to be of $Z$ and $X$ type, respectively, an assumption we make throughout. 
Let $\bl$ and $\bm$ be the measurement outcomes of all the logical operators $\{ \bar{Z}_j\ |\ 1 \leq j\leq n-m_x-m_z  \}$ and  $\{ \bar{X}_j\ |\ 1 \leq j\leq n-m_x-m_z  \}$, respectively, and $\bar{\Pi}_\bl:=\sum_{\m{k}\in [\bl]}P_\m{k}$ and $\hat{\Pi}_{\bm}:=\sum_{\m{x}\in [\bm]}\widetilde{P}_\m{x}$ the associated projectors for $[\bl]$ and $[\bm]$ the corresponding equivalence classes. 

The idea behind one-shot distillation is for Alice to measure the syndromes $\ba$ and $\bb$ 
on her system and reveal $\ba$ to Bob. If the CSS code is properly chosen, this information should make it possible to distinguish the corresponding marginals of his key system and the shield, at which point Theorem~\ref{TheoremMeasImplySecKey} would apply to key observables $\bar{Z}_j$ and conjugate observables $\bar{X}_j$. Bob only needs $\ba$, since the mere existence of the conjugate basis measurement implies the secrecy of the key. 
\jmr{In QKD, measuring the encoded $Z$ operators is equivalent to privacy amplification, and the degrees of freedom in defining the logical operators $\bar{Z}_j$ give rise to different families of privacy amplification functions. 
Here we present a one-shot private state distillation theorem useful for QKD security proofs~\footnote{The other major step in a complete proof is a parameter estimation scheme to determine what state Alice and Bob share, given their measurement results.}.} 

\begin{theorem}[One-Shot Distillation]
\label{th:oneshotdistillation}
\jmr{Let} Alice and Bob share an arbitrary state $\Psi^{ABS}$ with dim$(A)=d^n$ and purification $\ket{\Psi}^{ABSE}=\sum_\m{k}\sqrt{p_\m{k}}\ket{\m{k}}^A\ket{\varphi_\m{k}}^{BSE}$. Suppose there exists a CSS code with $m_z$ $Z$-type stabilizers and $m_x$ $X$-type stabilizers whose syndromes $\ba$ and $\bb$ \jmr{are associated with} measurements $\Lambda_{\ba,\m{k}}^B$ and $\widetilde{\Lambda}_{\bb,\m{x}}^{BS}$ for which  
\begin{align}
p_{{\rm e}}&= \sum_\ba \sum_{\m{j}\neq \m{k}}\Tr\left[(P^A_\m{j} \otimes \Lambda^B_{\ba,\m{k}})\Pi^A_\ba \Psi^{AB} \right]\leq \eps_{z},\\
\widetilde{p}_{{\rm e}}&=\sum_{\bb}\sum_{\m{x}\neq \m{y}}\Tr\left[(\widetilde{P}^A_\m{x} \otimes \widetilde{\Lambda}_{\bb,\m{y}}^{BS}) \widetilde{\Pi}^A_\bb \Psi^{ABS}  \right]\leq \eps_{x}.
\end{align}
\jmr{Then by one-way communication from Alice to Bob they can distill an $(\eps_{z}+\sqrt{\eps_x})$-private state of size $d^{n-m_z-m_x}$ whose key is the encoded value $\bl$.} 
\end{theorem}

\begin{proof}
Suppose that Alice measures the syndromes $\ba$ and $\bb$ and makes $\ba$ public. 
The post-measurement state is $\ket{\Psi_1}^{ABSERT}:=\sum_{\ba,\bb}\Pi^A_\ba \widetilde{\Pi}^A_\bb\ket{\Psi}^{ABSE}\ket{\ba}^{R}\ket{\bb}^T$ where $R$ is a new public register shared by all parties but $T$ is held by Alice.
Coherently measuring $\Lambda_{\ba,\m{k}}^B$ with the partial isometry $U^{BB_2R}$ produces 
\begin{align*}
	\ket{\Psi_2}:=U^{BB_2R}\ket{\Psi_1}=\sum_{\ba,\m{k}} \sqrt{\Lambda_{\ba,\m{k}}^B}\otimes P_\ba^R\ket{\Psi_1}^{ABSERT}\ket{\m{k}}^{B_2}.
\end{align*}
Bob can determine the values of $\bar{Z}^A_j$ for all $j$ with error probability
\begin{align*}
	p'_{\rm e}&=\sum_{\bl\neq \bl'}\Tr\left[\left(\bar{\Pi}_\ba^A\otimes \bar{\Pi}^{B_2}_{\bl'}\right)\Psi_2^{AB_2} \right]\\
&=\sum_{\bl\neq \bl'}\sum_{\ba,\bb}\sum_{\m{k}\in[\bl']}\Tr\left[\left(\bar{\Pi}_\bl^A\otimes \Lambda^B_{\ba,\m{k}}\right)\Pi^A_\ba\widetilde{\Pi}^A_\bb\Psi^{AB} \right]\\
&=\sum_{\bl\neq \bl'}\sum_{\ba}\sum_{\m{k}\in[\bl']} \Tr\left[\left(\bar{\Pi}_\bl^A\otimes \Lambda^{B}_{\ba,\m{k}}\right)\Pi^A_\ba\Psi^{AB} \right]\\
&\leq \sum_{\ba}\sum_{\m{j}\neq \m{k}}\Tr\left[\left(P_\m{j}^A\otimes \Lambda^{B}_{\ba,\m{k}}\right)\Pi^A_\ba\Psi^{AB} \right]\\
&\leq \eps_z,
\end{align*}
where we have used $[\bar{\Pi}^A_\bl, \widetilde{\Pi}^A_\bb ]=0$ and $\sum_{\bb} \widetilde{\Pi}^A_\bb = \mathbbm{1}^A$. Alice's conjugate basis measurement can be accurately predicted by first undoing $U^{BB_2R}$ and then 
measuring $\widetilde{\Lambda}^{BS}_{\bb,\m{y}}$. An entirely similar calculation shows that the resulting error probability
is less than $\eps_x$. Hence, by Theorem~\ref{TheoremMeasImplySecKey} $\Psi_2$ is an $(\eps_z+\sqrt{\eps_x})$-private state, 
whose key subsystems are the encoded subsystems $\bar{A}$ and $\bar{B}_2$. 
\end{proof}

As stated, the above theorem only involves one-way communication. However, it can easily be generalized to the sorts of two-way error-correction protocols presented in~\cite{gottesman_proof_2003}. The idea is that, instead of making only one measurement, Alice and Bob execute successive ``partial'' measurements of the syndrome of the dit error correction code, each of which is followed by a round of two-way classical communication. Each measurement is still associated with a set of $Z$-type operators, but the $Z$-type operators of the $i$th round of measurement could depend on all their previous outcomes. One-way error correction can be interpreted as the case in which the $Z$-type operators are chosen independently.

\subsection{Achievable distillation rates}

Now we turn to the achievable distillation rates. Define an $(n,\eps)$ distillation protocol for $\psi^{ABS}$ to be a series of local quantum operations and classical communication such that application on $\Psi^{ABS}=(\psi^{ABS})^{\otimes n}$ produces an $\eps$-private state. If there exists an $(n,\eps_n)$ protocol for every $n$, producing a $\log_2 \tau_n$-bit approximate private state, such that $\lim_{n\rightarrow \infty}\eps_n=0$, then the fractional yield of private outputs to raw inputs defines the achievable rate
\begin{align}
	R=\lim_{n\rightarrow \infty} \frac{\log_2 \tau_n}{n}.
\end{align}
Finally, the supremum of achieveable rates is called the one-way distillable privacy  $P_\rightarrow(\psi^{ABS})$ of the state $\psi^{ABS}$. 
In the following, we use the label ${\psi_a}$ where necessary to denote that the entropy or mutual information is computed using an extended version ${\psi_a}^{ACBSE}$ of the state $\psi^{ABSE}$. 
Using the previous result and a slightly modified version of the HSW theorem given in Appendix~\ref{app:HSW}, we quickly get the following: 
\begin{theorem}[One-Way Distillable Privacy]
\label{thm:dp}
Given conjugate observables $Z^A$ and $X^A$, consider an arbitrary state $\psi^{ABS}$ and its extension $\psi_a^{ACBS}$ obtained by copying the $Z^A$ basis of $A$ to $C$. Then
\begin{align*}
P_\rightarrow(\psi^{ABS})\geq\, &I(Z^A{:}B)-H(Z^A)+I(X^A{:}CBS)_{\psi_a}.
\end{align*}
\end{theorem}

\begin{proof}
Without loss of generality, we can assume
that $d={\rm dim}(A)$ is prime by appending additional $\ket{k}^A$ for which the corresponding weights $p_k=0$. 
Let $C$ be under Alice's control so that she can perform the copy operation and 
consider $\Psi_a^{ACBS}=(\psi_a^{ACBS})^{\otimes n}$. 
Pick a CSS code $c$ from the distribution $\mathcal{C}$ given in 
Appendix~\ref{ap:univCSS}, so that the $Z$-type and $X$-type stabilizers give rise to universal hash functions (for a definition, see Appendix~\ref{app:HSW}), and let $m_z=\frac{n}{\log_2 d}\left[H(Z^{A})-I(Z^{A}{:}B)+4 \delta\right]$ and 
$m_x=\frac{n}{\log_2 d}\left[H(X^{A})_{\psi_a}-I(X^{A}{:}CBS)_{\psi_a}+4 \delta\right]$ for a fixed $\delta>0$. 
Theorem~\ref{thm:HSW} implies that the measurements $\Lambda_{\ba,\m{k}}^B$ constructed 
from these hash functions can predict Alice's key with average error probability 
$\langle \eps_{z, c} \rangle_\mathcal{C} \leq 6 \cdot 2^{-n\delta^2}$. 
Similarly, the average error probability of 
the measurements $\widetilde{\Lambda}_{\bb,\m{x}}^{CBS}$ in predicting the conjugate basis observable is 
$\langle \eps_{x,c} \rangle_\mathcal{C}\leq 6 \cdot 2^{-n\delta^2}$.  Now apply Theorem~\ref{th:oneshotdistillation}
to each CSS code, where the shield is the combined system $CS$, and average over the different codes. 
Using the concavity of the square root and the fact that $H(X^{A})_{\psi_a}= \log_2 d$, it follows that 
Alice and Bob can create an $\eps$-private state having $n[I(Z^{A}{:}B)+I(X^{A}{:}CBS)_{\psi_a}-H(Z^{A})-8\delta]$ key bits, for $\eps\leq\langle \eps_{z, c} \rangle_\mathcal{C}+\sqrt{\langle \eps_{x, c} \rangle_\mathcal{C}} \leq 6 \cdot 2^{-n\delta^2} +\sqrt{6 \cdot 2^{-n\delta^2}}$. 
\end{proof}

By Lemma~\ref{lm:DWvsCOMPL}, $P_\rightarrow(\psi^{ABS})\jmr{\geq} I(Z^A{:}B)-I(Z^A{:}E)$, so this method achieves the same yield of secret key as the random coding method used by Devetak and Winter~\cite{devetak_distillation_2005}. 

\begin{lemma}
\label{lm:DWvsCOMPL}
For conjugate observables $Z^A$ and $X^A$ and a state of the form $\ket{\psi_a}^{ACBSE}=\sum_k \sqrt{p_k} \ket{k}^A\ket{k}^{C}\ket{\varphi_k}^{BSE}$, $I(X^{A}{:}CBS)=H(Z^{A})-I(Z^A{:}E)$.
\end{lemma}
\begin{proof}
Rewrite $\ket{\psi_a}^{ACBSE}$ as $\frac{1}{\sqrt{d}}\sum_x \cket{x}^A\ket{\vartheta_x}^{CBSE}$ for $\ket{\vartheta_x}^{CBSE}=Z_x^C\sum_k \sqrt{p_k}\ket{k}^{C}\ket{\varphi_k}^{BSE}$. Hence 
$S(\vartheta_x^{CBS})=S(\vartheta^{CBS}_0)$ for all $x$. From the Schmidt decomposition, $S(\vartheta^{CBS}_0)=S(\vartheta^{E}_0)=S(E)$ and  $S(CBS)=S(AE)$.  Therefore,  
\begin{align*}
I(X^A{:}CBS)&= S(CBS)- \sum_x q_x S(\vartheta^{CBS}_x)\\ &= S(AE)- S(\vartheta^{CBS}_0)\\ &= S\bigg(\sum_k p_k P_k^A\otimes \varphi_k^E\bigg)- S(E)\\ &= H(Z^A)-I(Z^A{:}E).\qedhere
\end{align*}\end{proof}

An immediate corollary is that the distillable privacy of 
an arbitrary state $\psi^{AB}$ without a specified shield system must be no less than the coherent information 
$ I_c(A\rangle B):= S(B)-S(AB)$; this can be seen as a weaker version of the hashing inequality, which we will consider
in the next section. 
\begin{corollary}
\label{cor:hasingkey}
$
P_\rightarrow(\psi^{AB})\geq I_c(A\rangle B).
$
\end{corollary}
\begin{proof}
Pick any observable $Z^A$ and define the computational basis of $A$ as its eigenbasis. 
Consider the purification $\ket{\psi}^{ABE}= \sum_k \sqrt{p_k}\ket{k}^A \ket{\varphi_k}^{BE}$ of $\psi^{AB}$, and note 
that $I_c(A \rangle B)= S(B)-S(E)= I(Z^A{:}B)-I(Z^A{:}E)$, where the last equality follows from the fact that $S(\varphi_k^B)=S(\varphi_k^E)$ for all $k$. From Theorem~\ref{thm:dp} and Lemma~\ref{lm:DWvsCOMPL}, $P_\rightarrow(\psi^{AB})\geq I(Z^A{:}B)-I(Z^A{:}E)=I_c(A \rangle B)$. 
\end{proof}

\section{Hashing Inequality}
\label{sec:HashIneq}
Now we turn to the related question of entanglement distillation and show how the above analysis can be modified
to prove the hashing inequality on the one-way distillable entanglement $E_\rightarrow(\psi^{AB})$, which is defined 
analogously to $P_\rightarrow(\psi^{ABS})$.  
There are two main differences with the methods used in the preceding section. The first is that for  
Theorem~\ref{thm:dp}, it does not matter how the shield is split between Alice and Bob, but of course for
entanglement distillation Alice and Bob must be able to locally untwist the private state. 
The difficulty comes from the first step, in which Alice copies her key to system $C$, which was then considered part of the shield. Here, we avoid this problem by showing that after Bob makes the $\Lambda_\ba^B$ measurement, he effectively has system $C$. Thus, he has the entire shield, and can perform the untwisting operator himself.

The second difference stems from the definition of approximate private states as states that yield approximate
secret keys when measured. Because we must now perform all measurements coherently, these results are not directly 
applicable. Modifying them is possible, but we prefer to give a more direct argument, which has the side benefit of 
yielding a better approximation parameter.

\begin{theorem}[Hashing Inequality]
\label{th:HashIneq}
$E_\rightarrow(\psi^{AB})\!\geq\! I_c(A\rangle B)$.
\end{theorem}

\begin{proof}
The proof proceeds by successively performing the $\Lambda^B_\ba$ and $\widetilde{\Lambda}^{B}_\bb$ measurements coherently and showing how the result is close to an entangled state.  
Purify $\psi^{AB}$ to $\ket{\psi}^{ABE}=\sum_{k=0}^{d-1} \sqrt{p_k} \ket{k}^A \ket{\varphi_k}^{BE}$. Without loss of generality, we can assume that $d={\rm dim}(A)$ is prime by appending additional states $\ket{k}$ for which $p_k=0$. 
Now define $\ket{\Psi}^{ABE}:=(\ket{\psi}^{ABE})^{\otimes n}=\sum_\m{k} \sqrt{p_\m{k}} \ket{\m{k}}^A \ket{\varphi_\m{k}}^{BE}$, 
where ${p}_\m{k}=p_{k_1}p_{k_2}\cdots p_{k_n}$.

Suppose Alice picks a CSS code $c$ from the distribution $\mathcal{C}$ described in Appendix~\ref{ap:univCSS} with $m_z$ $Z$-type and $m_x$ $X$-type stabilizers, measures the dit and phase error syndromes $\ba$ and $\bb$, and declares them publicly. 
This transforms the state into 
\begin{align}
	\ket{\Psi_1}:=\sum_{\ba,\bb}\Pi^A_\ba\widetilde{\Pi}^A_\bb\ket{\Psi}^{ABE}\ket{\ba,\bb}^R,
\end{align}
where $R$ is a publicly-held register. 

Let $m_z= \frac{n}{\log_2 {d}}\left[H(Z^{A})_\psi-I(Z^{A}{:}B)_\psi+4\delta\right]$ for some arbitrary $\delta >0$. By Theorem~\ref{thm:HSW}, there exists a measurement $\Lambda_{\ba}^B$ that predicts Alice's key with error probability $\eps_{z,c}$ such that $\langle\eps_{z, c} \rangle_\mathcal{C} \leq 6 \cdot 2^{-n \delta^2}$. Performing this measurement coherently yields 
\begin{align*}
	\ket{\Psi_2}:=\sum_{\m{k},\ba,\bb}\Pi^A_\ba\widetilde{\Pi}^A_\bb\sqrt{\Lambda^B_{\ba,\m{k}}}\ket{\Psi}^{ABE}\ket{\m{k}}^C\ket{\ba,\bb}^R,
\end{align*}
where the output is stored in system $C$. 
This state is essentially identical to the one in which Bob simply has a copy of Alice's key,
\begin{align}
	\ket{\Psi_2'}:=\sum_{\ba,\bb}\Pi^A_\ba\widetilde{\Pi}^A_\bb\ket{\Psi_a}^{ABCE}\ket{\ba,\bb}^R,
\end{align}
where $\ket{\Psi_a}=\ket{\psi_a}^{\otimes n}$, as defined in Theorem~\ref{thm:dp}, except that Bob holds $C$.  Computing the fidelity, we obtain
\begin{align*}
	\bracket{\Psi_2}{\Psi_2'}&=\sum_{\ba,\m{k}\in[\ba]}p_\m{k}\bra{\varphi_\m{k}}\sqrt{\Lambda_{\ba,\m{k}}^B}\ket{\varphi_\m{k}}^{BE}\\
&\geq \sum_{\ba,\m{k}\in[\ba]}p_\m{k}\bra{\varphi_\m{k}}\Lambda_{\ba,\m{k}}^B\ket{\varphi_\m{k}}^{BE}\geq 1-\eps_{z,c},
\end{align*}
using the fact that $\sqrt{\Lambda}\geq \Lambda$ for $0\leq \Lambda\leq \mathbbm{1}$. Since the fidelity bounds the 
trace distance via Tr$|\rho-\sigma|\leq 2\sqrt{1-F(\rho,\sigma)^2}$~\cite{fuchs_cryptographic_1999}, we have
Tr$|\Psi_2-\Psi_2'|\leq 2\sqrt{2\eps_{z,c}}$. 

Now rewrite $\ket{\Psi_2'}$ as $\ket{\Psi_2'}=\sum_\m{x}\sqrt{q_\m{x}}\cket{\m{x}}^A\ket{\vartheta_\m{x}}^{BCE}$ and 
let $m_x= \frac{n}{\log_2 {d}} \left[H(X^{A})_{\psi_a}-I(X^{A}{:}BC)_{\psi_a}+4\delta\right]$. By Theorem~\ref{thm:HSW}, there exists a measurement $\widetilde{\Lambda}_{\bb}^{BC}$ that can predict the outcome of a conjugate measurement on $A$ 
with  error probability $\eps_{x,c}$ such that $\langle\eps_{x, c} \rangle_\mathcal{C} \leq 6 \cdot 2^{-n \delta^2}$.
Starting from $\ket{\Psi_2'}$, suppose Bob coherently measures $\widetilde{\Lambda}_\bb$ and store the result in $D$. 
This gives
\begin{align*}
	\ket{\Psi_3'}:=\sum_{\m{y},\ba,\bb}\Pi^A_\ba\widetilde{\Pi}^A_\bb\sqrt{\widetilde{\Lambda}^{BC}_{\bb,\m{y}}}\ket{\Psi_a}^{ABCE}\cket{\m{y}}^D\ket{\ba,\bb}^R.
\end{align*}
As before, this is essentially the same as the state $\ket{\Psi_3''}$ in which Bob has a copy of Alice's string $\m{x}$ in 
system $D$,
\begin{align}
\label{eq:psi3pp}
\ket{\Psi_3''}=\sum_{\m{x},\ba,\bb}\sqrt{q_\m{x}}\,\Pi^A_\ba\widetilde{\Pi}^A_\bb\cket{\m{x}}^A\cket{\m{x}}^D\ket{\vartheta_\m{x}}^{BCE}\ket{\ba,\bb}^R,
\end{align}
and a similar calculation to the one above shows that Tr$|\Psi_3'-\Psi_3''|\leq  2\sqrt{2\eps_{x,c}}$.  

Implicit in rewriting $\ket{\Psi_2'}$ using Alice's conjugate basis is the fact that $\sqrt{q_\m{x}}\ket{\vartheta_\m{x}}^{BCE}=\sum_\m{k}\sqrt{p_\m{k}}\bracket{\widetilde{\m{x}}}{\m{k}}\ket{\m{k}}^{C}\ket{\varphi_\m{k}}^{BE}$. 
Substituting this in Eq.~(\ref{eq:psi3pp}) gives
\begin{align*}
	\ket{\Psi_3''}=&\frac{1}{\sqrt{d^n}}\sum_{\m{x},\ba,\bb}\Pi^A_\ba\widetilde{\Pi}^A_\bb\cket{\m{x}}^A\cket{\m{x}}^D\ket{\ba,\bb}^R\\
&\otimes \sum_\m{k}\sqrt{p_\m{k}}\,\omega^{{\m{x}\cdot\m{k}}}\ket{\m{k}}^C\ket{\varphi_\m{k}}^{BE}.
\end{align*}

Bob can now decouple subsystem $BCE$ by 
using the operator $U^{BD}=\sum_{\m{k},\m{x}}\omega^{-{\m{x}\cdot\m{k}}}\widetilde{P}^D_{\m{x}}\otimes P^B_\m{k}$,
and the result is an entangled state in the encoded subsystem $\bar{A}\bar{D}$,
\begin{align}
	\ket{\Psi_4''}:=U^{BD}\ket{\Psi_3''}=&\frac{1}{\sqrt{d^n}}\sum_{\ba,\bb}\Pi^A_\ba\widetilde{\Pi}^A_\bb\ket{\Phi_{d^n}}^{AD}\ket{\ba,\bb}^R\nonumber\\&\otimes\sum_\m{k}\sqrt{p_\m{k}}\ket{\m{k}}^C\ket{\varphi_\m{k}}^{BE}.
\end{align}

Since they never hold exactly $\ket{\Psi_2'}$ or $\ket{\Psi_3''}$, Alice and Bob only end up with a good approximation 
to an entangled state. To determine how good, we can use properties of the trace distance. Call 
the unitaries implementing the coherent measurements $U_z^{BC}$ and $U_x^{BCD}$, respectively, and define $W^{BCD}=U^{BD}U_x^{BCD}U_z^{BC}$. Applying $W$ to $\Psi_1$ generates $\Psi_4$, and by the triangle inequality and unitary invariance of the trace distance, we have 
\begin{align}
	\Tr\big|\Psi_4-\Psi_4''\big|\leq 2(\sqrt{2\eps_{z,c}}+\sqrt{2\eps_{x,c}}).
\end{align}
The next step is to average over all CSS codes. Using the concavity of the square root and the fact that 
the trace distance cannot increase under the partial trace, we obtain
\begin{eqnarray}
\Tr| {\Psi}_4^{\bar{A}\bar{D}} - \Phi^{\bar{A}\bar{D}}| \leq 8 \sqrt{3 \cdot 2^{-n \delta^2}}.
\end{eqnarray}

Finally, we must show that the resulting rate is given by the coherent information.  
Since $H(X^{A})_{\psi_a}= \log_2 {d}$, $(n-m_x-m_z) \log_2 {d}=  n \left[I(Z^{A}{:} B)_\psi+I(X^{A}{:}BC)_{\psi_a}-H(Z^{A})_\psi-8\delta\right]$. By Lemma~\ref{lm:DWvsCOMPL}, $I(X^{A}{:}BC)_{\psi_a}=H(Z^A)_{\psi_a}-I(Z^{A}{:}E)_{\psi_a}$. Clearly $H(Z^A)_\psi=H(Z^A)_{\psi_a}$ and similarly for the quantum mutual information of $Z^A$ with $B$ or $E$.  Since $I(A\rangle B)_{\psi_a}=I(Z^{A}{:}B)_\psi-I(Z^{A}{:}E)_\psi$, as in Corollary~\ref{cor:hasingkey}, $(n-m_x-m_z) \log_2 {d}= nI_c(A\rangle B)_\psi-8n \delta$, which concludes the proof. 
\end{proof}

\section{Relation to previous work}
\label{sec:previouswork}

The present work is an outgrowth of earlier work on private states by one of us~\cite{renes_noisy_2007} and draws much inspiration from the work of Koashi~\cite{koashi_unconditional_2006,koashi_complementarity_2007}. In particular, Theorems~\ref{TheoremMeasImplySecKey} and~\ref{thm:goodkcmexist} are closely related to the first two theorems of~\cite{koashi_complementarity_2007}, in which Koashi defines the two protocols of the complementary control scenario. It is easy to see that our condition on the predictability of the key is equivalent to his condition on the primary protocol, and that our condition on the measurement $\widetilde{\Lambda}^{BS}$ implies his condition on the secondary protocol. Therefore, Theorem~\ref{TheoremMeasImplySecKey} is a corollary of the first theorem of~\cite{koashi_complementarity_2007}. Although we were not able to show that the condition on the secondary protocol implies our condition on the measurement $\widetilde{\Lambda}^{BS}$, Theorem~\ref{thm:goodkcmexist} can be proven using arguments very similar to those found in~\cite{koashi_complementarity_2007}.

Meanwhile, Theorem~\ref{th:oneshotdistillation} corresponds conceptually to the inclusion of the complementary control scenario in the security analysis of~\cite{koashi_unconditional_2006}, with several important differences in the 
details. First, we do not consider parameter estimation at all, while~\cite{koashi_unconditional_2006} presents 
a full security analysis for BB84. To complete a security proof using our results, one would need to determine
what quantum states $\psi^{ABS}$ are compatible with the output of the parameter estimation 
phase of the protocol in order to apply Theorems~\ref{th:oneshotdistillation} and~\ref{thm:dp}. This can be done with an estimate of the quantum channel noise obtained indirectly from the experimental measurements. The advantage of Theorem~\ref{th:oneshotdistillation} is that it could be used to prove the security of a more general set of QKD protocols, even those
including preprocessing. 
Second, \cite{koashi_unconditional_2006} assumes that Bob's conjugate measurement is independent of $\bb$, with the supplemental information supplied only after the measurement is made. In our method, Bob uses the syndrome $\bb$ to construct the measurement $\widetilde{\Lambda}_{\bb}^{BS}$. 
Generally, the latter is no less powerful than the former, and avoids the pitfalls of \emph{locking} of accessible information~\cite{divincenzo_locking_2004}. In Appendix~\ref{app:ontheoneshot} we provide a concrete example in which allowing $\widetilde{\Lambda}^{BS}_\bb$ to depend on $\bb$ yields a better security parameter than if it were independent. 

The smaller difference concerns the step in \cite{koashi_unconditional_2006} of having Alice 
encrypt the amplitude error syndromes using a preshared secret key. This removes the need to use a CSS code~\cite{lo_method_2003}, but requires a key of size $O(n\log d)$ bits [in addition to the authentication key, of size $O(\log n\cdot\log d)$] and makes a small but practically significant difference for QKD. 
Theorem~\ref{th:oneshotdistillation} can be modified to encrypt the syndrome $\ba$ of an arbitrary (not necessarily linear) code as follows.
Supposing Alice and Bob already share a perfect secret key
$\boldsymbol{\ell}$ of the same size as the amplitude error syndrome $\ba$. Alice publicly transmits
$\ba+\boldsymbol{\ell}$ to Bob. He recovers $\ba$ using $\boldsymbol{\ell}$ and
can then make the $\Lambda_\ba^B$ measurement. 
The system $R$ storing the value of $\ba$ is unknown to Eve and 
can be decoupled with the operator 
$\sum_{\ba} \Pi^B_\ba \otimes (X^R)^{-\ba}$ since this does not affect
the key measurements. We can now apply Theorem~\ref{th:oneshotdistillation}
directly on the resulting correlated state. Using these ideas, one can easily show that the final
security parameter would have a similar form with or without
encrypting of the dit error syndrome.

By adapting Koashi's complementarity scenario, we are able to construct a means for distilling private states from arbitrary resource states at a rate given by the quantum Csisz\'ar-K\"orner bound. This complements the result of Devetak and Winter~\cite{devetak_relating_2004}, showing more directly how physical (quantum-mechanical) phenomena are responsible for the privacy of the key. As mentioned before, it must be possible to view their result as private state distillation by performing the operations coherently, and indeed a twisting operator plays an important role in their derivation of the hashing inequality, specifically the operator $U$ defined on p.~8 of~\cite{devetak_distillation_2005}. 
Mathematically speaking, the difference in the two approaches can be traced to the origins of this operator: here from the measurement used in the HSW theorem to determine the outcome of Alice's conjugate measurement, there from the quantum Chernoff bound via Uhlmann's theorem. 

A different approach to private state distillation is taken in~\cite{horodecki_quantum_2006}, whose ultimate goal is 
to show that key distribution is still possible over channels whose quantum capacity is zero, rather
than give rates on private state distillation. The distillation portion 
of the protocol accepts only certain inputs, namely twisted versions of noisy entangled states, and thus the distillation procedure works by untwisting the state and then applying entanglement distillation. The difficulty in this scheme 
then lies in determining the optimal combination of twisting operator and noise such that the given input can
be expressed in this form. As such, no closed-form distillation rate expressions can be given, and happily this is
not relevant to their goal. 

Our method of private state distillation gives a new proof of the hashing inequality, which then also implies a new proof of the direct quantum coding theorem. This version differs from previous work~\cite{lloyd_capacity_1997,shor_quantum_2002,devetak_private_2005, horodecki_quantum_2007, devetak_distillation_2005, hayden_decoupling_2007, hayden_random_2007} 
in several ways, mainly by the explicit use of CSS codes from the beginning and the fact that the decoder is constructed from the measurement used in the HSW theorem, rather than by decoupling Eve and appealing to Uhlmann's theorem. This construction resolves the open question raised in the conclusion of~\cite{hayden_random_2007} as here the decoder is directly linked to the bit and phase syndromes of the CSS code.

Finally, we would like to point out the connections to recent work on complementary channels. In~\cite{kretschmann_information-disturbance_2006,kretschmann_continuity_2007,kretschmann_complementarity_2007}, it has been shown that a correctable channel implies that the complementary channel is private, and vice versa. Theorems~\ref{TheoremMeasImplySecKey} and~\ref{thm:goodkcmexist} are essentially a static version of this (dynamic) result, applied to 
bipartite states instead of channels and starting from different assumptions. 

\section{Conclusion}
\label{sec:concl}

We provide a characterization of private states in terms of \jmr{a complementary information tradeoff} and generalize the security proof methods based on entanglement distillation and the uncertainty principle. This generalization is formulated as a one-shot distillation theorem (Theorem~\ref{th:oneshotdistillation}). Exploiting this framework, we give alternative proofs of the quantum Csisz\'ar-K\"orner bound on distillable secret key (Theorem~\ref{thm:dp} and Lemma~\ref{lm:DWvsCOMPL}) and the hashing inequality on distillable entanglement (Theorem~\ref{th:HashIneq}).

One of the main applications of this work is of course to QKD, particularly proofs for realistic protocols. These involve more physical systems than just those describing the keys and the eavesdropper, and one challenge has been determining how to use information the honest parties have about such systems. Including the shield system into the security analysis and picturing the QKD process as private state distillation gives a general method for doing so, a point also emphasized by Koashi~\cite{koashi_unconditional_2006}. The importance of these extra systems is how they contribute to knowledge of hypothetical conjugate basis measurements made on the key system of either party. 

This is dramatically exemplified by Koashi's security proof of the BB84 protocol with uncharacterized detectors, which proceeds by noting that this protocol directly furnishes Bob with an estimate of Alice's conjugate basis result, regardless of the detector details. Our results provide a more detailed and complete picture of how shield systems contribute to privacy, which should expand the range of protocol and device imperfections that can be treated. For instance,
it would be interesting to investigate the unconditional  security of QKD protocols that are not permutation invariant~\cite{inoue_differential_2002, stucki_fast_2005}. This possibility is particularly appealing since Theorem~\ref{th:oneshotdistillation} does not require a permutation of the input state nor does it depend on a particular method of parameter estimation. We plan to examine these issues and other implications for realistic protocols in an upcoming publication. 

As a final remark, we note that our approach to the hashing inequality is closely related to~\cite{hayden_random_2007}, which also makes use of an information-uncertainty relation. In fact, that relation is simply the ``quantum'' version of the complementary information tradeoff, Lemma~\ref{lemma:cit}, replacing the classical conditional entropy $H$ with the classical-quantum conditional entropy $S$ to obtain
\begin{equation}
S({Z}^A|E)+S(\widetilde{X}^A|B)\geq \log_2 d
\end{equation}
for any state $\rho^{ABE}$, conjugate observables $Z^A$ and $\widetilde{X}^A$, and $d={\rm dim}(A)$. As the ``classical'' version
can easily be generalized to nonconjugate observables simply by using the general form of the entropic uncertainty relation, it becomes reasonable to ask if the ``quantum'' version of the same holds as it does for strictly conjugate observables. Numerical evidence supports this claim, and we explore this subject in more detail in~\cite{renes_strong_2008}.

\section*{ACKNOWLEDGEMENTS}
We thank Gernot Alber, Aram Harrow, Hoi-Kwong Lo, Norbert L\"utkenhaus, and
Graeme Smith for helpful discussions. 
J.M.R.\ received support from the Alexander von Humboldt Foundation and the 
European IST project SECOQC, and J.-C.B.\ from the
Natural Sciences and Engineering Research Council of Canada and Quantumworks.

\appendix

\section{APPROXIMATE PRIVATE STATE PROOFS}
\label{app:proofs}
Here we present the proofs of Theorems~\ref{TheoremMeasImplySecKey} and \ref{thm:goodkcmexist}.

\begin{proof}[Proof of Theorem~\ref{TheoremMeasImplySecKey}]
	
Write the purification of $\psi^{ABS}$ as $\ket{\psi}^{ABSE}=\sum_{jk}\sqrt{p_{jk}}\ket{jk}^{AB}\ket{\varphi_{jk}}^{SE}$ for some (normalized) states $\ket{\varphi_{jk}}^{SE}$. Copying the standard basis of Bob's state to a blank register $\ket{0}^{B'}$ with the unitary $C^{BB'}$ yields $\ket{\psi_1}^{ABSEB'}=\sum_{jk}\sqrt{p_{jk}}\ket{jk}^{AB}\ket{k}^{B'}\ket{\varphi_{jk}}^{SE}$. Let $\bar{\psi}^{ABB'SE}_1$ be the state after measuring $Z^A$ and $Z^B$ and consider the related state $\ket{\psi'_1}^{ABB'SE}=\sum_k\sqrt{p_{jk}}\ket{jj}^{AB}\ket{k}^{B'}\ket{\varphi_{jk}}^{SE}$. Performing the same measurement on $\psi'$ and computing the trace distance between the states, we find
\begin{align}
\Tr|\bar{\psi}_1^{ABE}-\bar{\psi}'^{ABE}_1|=2\sum_{j\neq k}p_{jk}=2p_{\rm e}\leq 2\eps_z.
\end{align}
Observe that $\ket{\psi_1'}^{ABB'SE}=C^{AB}\ket{\psi}^{AB'SE}\ket{0}^B$. Rewrite the original state as $\ket{\psi}^{AB'SE}=\sum_{x}\sqrt{q_x}\cket{x}^A\ket{\vartheta_x}^{B'SE}$ for some probability distribution $q_x$ and normalized states $\ket{\vartheta_x}^{B'SE}$. Coherently performing the $\widetilde{\Lambda}^{B'S}_y$ measurement with unitary $U^{B'ST}$, where the extra system $T$ stores the result, we find 
\begin{align}
	\ket{\psi_2}&=C^{AB}U^{B'ST}\ket{\psi}^{AB'SE}\ket{0}^B\ket{0}^T\\
&=\sum_{xy}\sqrt{q_x}C^{AB}\cket{x}^{A}\ket{0}^{B}\sqrt{\widetilde{\Lambda}_y^{B'S}}\ket{\vartheta_x}^{B'SE}\ket{y}^T.
\end{align}
Define $\ket{\psi_2'}=\sum_x\frac{\sqrt{q_x}}{\sqrt{1-\widetilde{p}_{\rm e}}}C^{AB}\cket{x}^{A}\ket{0}^B\sqrt{\widetilde{\Lambda}_x^{B'S}}\ket{\vartheta_x}^{B'SE}\ket{x}^T$; its fidelity with  $\ket{\psi_2}^{AB'SET}$ is
\begin{equation}
\bracket{\psi_2}{\psi_2'}=\sqrt{1-\widetilde{p}_{\rm e}}\geq \sqrt{1-\eps_x}.
\end{equation}

 In general, the fidelity between two quantum states is defined as $F(\rho,\sigma):=\Tr |\sqrt{\rho}\sqrt{\sigma}|$.
 Note that $\ket{\psi_2'}^{ABB'SET}$ is a private state with key systems $AB$ and shield $B'ST$. One way to see
this is to rewrite $\cket{x}$ in terms of $\ket{k}$,
\begin{equation*}
\ket{\psi_2'}=\frac{1}{\sqrt{d}}\sum_{kx}\frac{\sqrt{q_x}}{\sqrt{1-\widetilde{p}_{\rm e}}} e^{i\theta_{kx}}\ket{kk}^{AB}\sqrt{\Lambda_x^{B'S}}\ket{\vartheta_x}^{B'SE}\ket{x}^T.
\end{equation*}
Applying the unitary operator $W^{BT}=\sum_{kx}e^{-i\theta_{kx}}P_k^B\otimes P_x^T$ results in 
a maximally entangled state $\ket{\Phi}^{AB}$ in the $AB$ subsystem. Since $W^{BT}$ is a twisting operator, 
$\ket{\psi_2'}$ is a private state. 

If we now define $\ket{\psi_3}^{ABB'SET}=U^{\dagger B'ST}\ket{\psi_2'}^{ABB'SET}$, also a private state since $U^{\dagger B'ST}$ acts only on the shield, it follows from unitary invariance of the inner product that
\begin{equation}
F\left(\ket{\psi_3}^{ABB'SET},\ket{\psi_1'}^{ABB'SE}\ket{0}^T\right)\geq \sqrt{1-\eps_x}.
\end{equation}
Finally, bound the trace distance with the fidelity, using the relation $\Tr| \rho - \sigma| \leq \sqrt{1-F(\rho,\sigma)^2}$. This implies $\Tr|\bar{\psi}_3^{ABE}-\bar{\psi}_1'^{ABE}| \leq 2\sqrt{\eps_x}$, and using the triangle inequality we obtain $\Tr |\bar{\psi}^{ABE}-\bar{\psi}_3^{ABE}|\leq 2(\eps_z+\sqrt{\eps_x})$.
\end{proof}

\begin{proof}[Proof of Theorem~\ref{thm:goodkcmexist}]
Assume Eve holds the purification of $\psi^{ABS}$ and measure $AB$ to create the key. This yields $\bar{\psi}^{ABE}=\sum_{jk}(P^A_j\otimes P_k^B)\psi^{ABE}(P^A_j\otimes P_k^B)$.
A simple and direct calculation using the triangle inequality gives $2p_{\rm e} \leq \Tr|\bar{\psi}^{AB}-\kappa^{AB}|$. Since $\psi^{ABS}$ is an $\epsilon$-approximate private state, $\Tr |\bar{\psi}^{ABE}-\kappa^{ABE}| \leq 2\eps$. Tracing out $E$ does increase this distance, therefore $p_{\rm e}\leq \eps$.

To prove the analogue statement for the conjugate basis, we must define a suitable $\widetilde{\Lambda}^{BS}$. For this we adapt the corresponding measurement from the purification of $\kappa^{ABE}$, which is a private state. First bound the fidelity with the trace distance, using the fact that $1-\frac{1}{2}\Tr |\rho-\sigma|\leq F(\rho,\sigma)$~\cite{fuchs_cryptographic_1999}. Thus $F(\bar{\psi}^{ABE},\kappa^{ABE})\geq 1-\eps$. Uhlmann's theorem asserts that for any purification $\ket{{\psi}}^{ABER}$ of $\bar{\psi}^{ABE}$, there exists a purification $\ket{\kappa}^{ABER}$ of $\kappa^{ABE}$ such that $F(\bar{\psi}^{ABE},\kappa^{ABE})=F(\ket{\psi}^{ABER}, \ket{\kappa}^{ABER})$. We can set $R=SA'B'$ and take the former purification to be $\ket{\psi}^{ABER}:=C^{AA'}C^{BB'}\ket{\psi}^{ABSE}\ket{0}^{A'}\ket{0}^{B'}$ for $C^{AA'}$ and $C^{BB'}$ unitary operations such that $C^{AA'}\ket{k}^A\ket{0}^{A'}=\ket{k}^{A}\ket{k}^{A'}$. 

By definition, $\ket{\kappa}^{ABER}$ is an exact private state, and so is $\ket{\kappa^\prime}^{ABER}:=C^{\dagger AA'}C^{\dagger BB'}\ket{\kappa_p}^{ABER}$. Since fidelity is invariant under a unitary transformation, $F(\ket{\psi}^{ABSE}\ket{0}^{A'}\ket{0}^{B'}, \ket{\kappa^\prime}^{ABER})=F(\ket{\psi}^{ABER}, \ket{\kappa}^{ABER})$. Hence there exists $\Lambda^{\prime BR}_y$ such that measuring $\widetilde{P}^A_x\otimes \Lambda^{\prime BR}_y$ on $\ket{\kappa^\prime}^{ABER}$ produces the uniform distribution $\frac{1}{d}\delta_{xy}$. 
Making the same measurement on $\ket{\psi}^{ABSE}\ket{0}^{A^\prime}\ket{0}^{B^\prime}$ results in some probability distribution $\widetilde{q}_{xy}$. Observe that measuring $\Lambda^{\prime BR}_y$ on $\ket{\psi}^{ABSE}\ket{0}^{A^\prime}\ket{0}^{B^\prime}$ is the same as measuring $\Lambda^{BS}_y := \bra{00}^{A'B'}\Lambda'^{BR}_y\ket{00}^{A'B'}$ on $\ket{\psi}^{ABSE}$.

Since a quantum operation cannot decrease the fidelity, we immediately have $F(\ket{\psi}^{ABSE}\ket{0}^{A'}\ket{0}^{B'}, \ket{\kappa^\prime}^{ABER})\leq F(\widetilde{q}_{xy},\frac{1}{d}\delta_{xy})$. But 
\begin{equation}
	F\left(\widetilde{q}_{xy},\frac{1}{d}\delta_{xy}\right)=\frac{1}{\sqrt{d}}\sum_x \sqrt{\widetilde{q}_{xx}}\leq \sqrt{\sum_{x\neq y}\widetilde{q}_{xy}}=\sqrt{1-\widetilde{p}_{\rm e}} 
\end{equation}
by the concavity of the square root function. Collecting the inequalities, we find $\widetilde{p}_{\rm e}\leq 2\eps-\eps^2$.
\end{proof}

\section{STATIC HSW THEOREM}
\label{app:HSW}

Suppose a source described by the ensemble $\mathcal{E}=\{p_k,\varphi_k\}$ distributes classical letters $k\in\{0,1,\dots, d{-}1\}$ to Alice and quantum states $\varphi_k$ to Bob. Alice would like to communicate the value of $k$ to Bob, using as few resources as possible. Bob already possesses some information about $k$ in the form of $\varphi_k$, but in general cannot reliably distinguish between all these states. But Bob can learn $k$ if Alice reveals some information about $k$, a ``hint'' that narrows the set of $\varphi_k$ to some that he can reliably distinguish. 

This is the ``static'' version, first studied in~\cite{winter_coding_1999, devetak_classical_2003}, of the standard HSW scenario in which Alice actively encodes the information $s$ she wants to send to Bob using the signal ensemble $\mathcal{E}$~\cite{holevo_capacity_1998,schumacher_sending_1997}. Typically this problem is considered in the asymptotic setting of many identical and independent samples from $\mathcal{E}$. Alice then encodes her information into a block of such samples and Bob performs a collective measurement, a version of the so-called pretty good measurement (PGM)~\cite{hausladen_pretty_1994}, to decode the message. Properties of typical sequences and subspaces are used to prove that the PGM has a low probability of error.

Although in the main text we are concerned with using linear functions to generate the side information, in this appendix we shall consider the more general method of \emph{universal hashing}~\cite{carter_universal_1979} (also called 2-universal hashing), since it is not any more difficult and random linear functions are universal. In universal hashing the hint is generated by choosing a random $f:\{0,\dots, d^n{-}1\}\rightarrow \{0,\dots, m{-}1\}$ from a family $\mathcal{F}$ of hash functions and computing $t=f(x)$. Each function defines the subset  $\mathcal{S}_{t}$ of possible inputs having the same output value; hopefully Bob will be able to distinguish between the elements of this set. The family is called universal when the probability of collision, $f(x)=f(y)$ for $x\neq y$, is the same as for random functions: Pr$_f[f(x)=f(y)]\leq 1/m$. Put differently, the probability of any two elements being included in some $\mathcal{S}_t$ is also the same as if Alice chose the subsets completely at random, which is random enough for the procedure to work. 

In the i.i.d.~scenario Alice and Bob share $n$ copies of the state $\psi^{AB}=\sum_{k=0}^{d-1} p_k P_k^A\otimes \varphi_k^B$, which we 
write as $\Psi^{AB}=\sum_{\m{k}} p_\m{k}P_\m{k}^A\otimes\varphi_\m{k}^B$. 
By the following static HSW theorem, a hint roughly of size $\log_2 m=n\left[H(p_k)-\chi(p_k,\varphi_k)\right]=n\left[H(Z^A)-I(Z^A{:}B)\right]$ suffices for Bob to learn $\m{k}$ with exponentially small average probability of error.
\begin{theorem}[Static HSW Theorem for Universal Hash Functions]
\label{thm:HSW}
For $n$ copies of an arbitrary state of the form $\psi^{AB}=\sum_{k=0}^{d-1} p_k P_k^A\otimes \varphi_k^B$, fix $\delta>0$. 
Then for a universal family of hash functions $f: \{0, \dots, d^n-1\} \rightarrow \{0, \dots, m-1 \}$ where $\log_2 m=n\left[H(Z^A)-I(Z^A{:}B)+4\delta\right]$ there exist measurements $\Lambda_{f(\m{k}),\boldsymbol{\ell}}$ such that
\begin{align}
p_{\rm e}= \left\langle \sum_{\boldsymbol{\ell} \neq \m{k}}  \Tr\left[\Lambda_{f(\m{k}), \boldsymbol{\ell}}\,\varphi_{\m{k}}\right] \right\rangle_{f, \m{k}} \leq 6\cdot 2^{-n\delta^2}.
\end{align}
\end{theorem}
\begin{proof}
Fix a $\delta>0$ and start by Alice measuring her share of the state in the computational basis. With
probability greater than $1-\eps$ for $\eps=e^{-\frac{n\delta^2}{2}}$, the resulting string $\m{k}$ is typical, 
meaning  $\m{k}\in \mathcal{T}_\delta^n=\{\boldsymbol{\ell}|2^{-nH(p_k)-n\delta} \leq p_{\boldsymbol{\ell}} \leq  2^{-nH(p_k)+n\delta}$\}~\cite{cover_elements_1991}. If $\m{k}$ is not typical, the protocol aborts. 

If it does not abort, Alice randomly picks $f$ from a universal family $\mathcal{F}$ and sends $f(\m{k})$ to Bob via the public channel. 
This narrows the set of possible $\m{k}$ to the subset $\mathcal{C}_{f(\m{k})}$ of typical elements of 
$\mathcal{S}_{f(\m{k})}$. Bob will try to determine $\m{k}$ by making a measurement to distinguish the $\varphi_{\boldsymbol{\ell}}$ for $\boldsymbol{\ell}\in\mathcal{C}_{f(\m{k})}$. For this he uses the PGM defined by Eq.~(11) in~\cite{holevo_capacity_1998}, which is represented by the 
POVM elements 
 \begin{eqnarray*} 
 \Lambda^B_{f(\m{k}),\boldsymbol{\ell}}= \bigg(\sum_{\boldsymbol{\ell}\in\mathcal{C}_{f(\m{k})}}Q Q_{{\boldsymbol{\ell}}}Q\bigg)^{-\frac{1}{2}} Q Q_{{\m{k}}}Q \bigg(\sum_{\boldsymbol{\ell}\in\mathcal{C}_{f(\m{k})}}Q Q_{{\boldsymbol{\ell}}}Q\bigg)^{-\frac{1}{2}},
 \end{eqnarray*} 
where $Q$ and $Q_{\m{k}}$ are the projections into the typical subspaces (subspaces spanned by eigenstates with typical eigenvalues) of 
$\bar{\varphi}^{\otimes n}$ and $\varphi_{{\m{k}}}$, respectively. For a specific $f$ and $\m{k}$, a bound for the average error probability of this measurement is given by Eq.~(17) of~\cite{holevo_capacity_1998}, except that we do not yet need to average over all codewords,
 \begin{eqnarray*}
p_{\rm e}(\m{k})  &\leq& 3\Tr[\varphi_{\m{k}}(\mathbbm{1}-Q)] +\Tr[\varphi_{\m{k}}(\mathbbm{1}-Q_{\m{k}})]\\&&+\sum_{\boldsymbol{\ell}\in\mathcal{C}_{f(\m{k})}}\Tr[Q\varphi_{\m{k}} QQ_{\boldsymbol{\ell} }] +\eta_{\m{k}}, 
 \end{eqnarray*}
where $\eta_{\m{k}}$ is 1 if $\m{k}$ is typical and 0 otherwise. In our case, we are interested in the probability of error averaged over all $f$ and $\m{k}$, i.e.~$\langle P_{\rm e}(\m{k})\rangle_{f, \m{k}}$. To compute it, we need the following relations (see \cite{holevo_capacity_1998} for details): 
\begin{align}
\Tr[\bar{\varphi}^{\otimes n}(\mathbbm{1}-Q)] &\leqslant \epsilon,\\
\left\langle\Tr[\varphi_{\m{k}}(\mathbbm{1}-Q_{\m{k}})]\right\rangle_{\bf k} &\leqslant \epsilon,\\
Q_{\m{k}} &\leqslant 2^{n\sum_ip_iS(\varphi_i)+n\delta}  \varphi_{{\m{k}}},\\
\sum_{\m{k}\in\mathcal{T}_\delta^n} \varphi_{{\m{k}}} &\leqslant 2^{nH(p_i)+n\delta} \bar{\varphi}^{\otimes n},\\
||Q\bar{\varphi}^{\otimes n} Q||_\infty &\leqslant 2^{-nS(\bar{\varphi})+n\delta}, \end{align} where $||M||_\infty$ is the maximal eigenvalue of $M$. Since $\langle \varphi_{\m{k}}\rangle_{\m{k}}=\bar{\varphi}^{\otimes n}$, we have
   \begin{eqnarray*}
\langle P_{\rm e}(\m{k})\rangle_{\m{k}, f} &\leq& 5\epsilon +\langle \sum_{\bm \in\mathcal{C}_{f(\m{k})}}\Tr[Q\varphi_{\m{k}} QQ_{\boldsymbol{\mu}}] \rangle_{\m{k}, f}\\
&\leq& 5\epsilon+ \langle \sum_{\boldsymbol{\mu}\in\mathcal{T}_\delta^n} {\rm Pr}_f[f(\boldsymbol{\mu})=f(\m{k})]{\rm Tr}[Q\varphi_{\m{k}} QQ_{\bm}]\rangle_{\m{k}}.
\end{eqnarray*}
Straightforward calculations give
 \begin{eqnarray*}
 \langle P_{\rm e}(\m{k})\rangle_{\m{k}, f} &\leq& 5\epsilon+\frac{1}{m}{2^{nH(p_i)+n\sum_ip_iS(\varphi_i)+2n\delta}}\,\Tr[Q\bar{\varphi}^{\otimes n}Q \bar{\varphi}^{\otimes n}]\\ 
 &\leq&5\epsilon+\frac{1}{m}{2^{nH(p_i)-nS(\bar{\varphi})+n\sum_ip_iS(\varphi_i)+3n\delta}}, 
\end{eqnarray*}
where for the last step we use the relation $\Tr[Q\bar{\varphi}^{\otimes n} Q\bar{\varphi}^{\otimes n}] \leqslant ||Q\bar{\varphi}^{\otimes n} Q||_\infty \Tr[\bar{\varphi}^{\otimes n}]= ||Q\bar{\varphi}^{\otimes n} Q||_\infty$. Choosing $ \log_2 m\geq n\left[H(p_i)-S(\bar{\varphi})+\sum_ip_iS(\varphi_i)+4\delta\right]$ completes the proof.
\end{proof}

\section{UNIVERSAL DISTRIBUTION FOR STABILIZERS OF CSS CODES}
\label{ap:univCSS}

The question we answer in this section is how to pick a family of CSS codes such that both the $Z$- and $X$-type stabilizers are universal hash functions. The difficulty is that the two stabilizers are not independent; they must commute with each other. 
The $Z$ and $X$ stabilizers can be represented by an $m_z$ by $n$ matrix $M_z$ and the  $m_x$ by $n$ matrix $M_x$, respectively, where each entry is an integer modulo $d$. We have the following

\begin{lemma}
Consider the set of all $m_x+m_z$ by $n$ matrices $R$ such that each row is orthogonal to the others and where each entry is an integer modulo a prime number $d$. Let $M_z$ be the first $m_z$ rows of $R$, and $M_x$ be the last $m_x$ rows of $R$. Then the linear functions associated with $M_z$ and $M_x$ are both universal.
\end{lemma}

\begin{proof}
Let $\m{r}_{i}$ be the $i$th row of $R$. All possible strings have the same probability to be $\m{r}_1$. 
Therefore, for any distinct $n$ dit-strings $\m{k}$ and $\m{k}'$, ${\rm Pr}_R[\m{r}_1 \cdot \m{k}=\m{r}_1 \cdot \m{k}'] = \frac{1}{d}$. This is not generally true if $d$ is not prime. Now we proceed by induction. Assume that we have a set $R_{\ell}$ of strings $\m{r}_1$, $\m{r}_2$, ... and $\m{r}_\ell$ such that ${\rm Pr}_R[\m{r}_i \cdot \m{k}=\m{r}_i \cdot \m{k}'\ |\ 1 \leq i \leq \ell ] \leq \frac{1}{d^\ell}$. 
Conditional on $R_\ell$, the next row $\m{r}_{\ell+1}$ is uniformly distributed over the space of strings orthogonal to the set $R_\ell$.  If $\m{r}_j \cdot \m{k} \neq \m{r}_j \cdot \m{k}'$ for some $1 \leq j \leq \ell$, then ${\rm Pr}[\m{r}_i \cdot \m{k}=\m{r}_i \cdot \m{k}'\ |\ 1 \leq i \leq \ell+1 ]=0$. So we can consider only the case in which $\m{r}_i \cdot \m{k} = \m{r}_i \cdot \m{k}'$ for all $1 \leq i \leq \ell$. In that situation, $\m{k}-\m{k}'$ can be expended in any basis of the space orthogonal to $R_\ell$ (the coefficients being integers from $0$ to $d-1$). Pick one such basis. $\m{r}_{\ell+1}$ is uniformly distributed over all strings that are spanned by this basis, therefore ${\rm Pr}_{R | R_\ell}[\m{r}_{\ell+1}\cdot \m{k}=\m{r}_{\ell+1}\ \cdot \m{k}' ]=\frac{1}{d}$, where we assumed $\m{r}_i \cdot \m{k} = \m{r}_i \cdot \m{k}'$ for all $1 \leq i \leq \ell$. Including all possible cases, we deduce that ${\rm Pr}_{R}[\m{r}_{i}\cdot \m{k}=\m{r}_i \cdot \m{k}'\ |\ 1 \leq i \leq \ell+1 ]<\frac{1}{d^{\ell+1}}$.

Since there is no distinction between the order of the rows of $R$, we conclude that any function associated with a matrix composed of a subset of rows of $R$ is universal.
\end{proof}

\section{ON THE ONE-SHOT DISTILLATION THEOREM}
\label{app:ontheoneshot}

Parameter estimation aside, Theorem~\ref{th:oneshotdistillation} is stronger than the security proof of~\cite{koashi_unconditional_2006}. Constructing an example where this is the case is not too difficult and we will simply give an example in which the optimal $\widetilde{\Lambda}^{BS}_{\beta}$ for guessing Alice's conjugate basis measurement is not independent of $\beta$. Consider two copies (i.e.~$n=2$) of the state 
\begin{align*}
\ket{\psi}^{ABSE}&=\frac{1}{2}(\ket{0}^A\ket{0}^B+\ket{1}^A\ket{1}^B)\ket{\phi_0}^S \ket{0}^E \\&+\frac{1}{2}(\ket{0}^A\ket{0}^B-\ket{1}^A\ket{1}^B)\ket{\phi_1}^S \ket{1}^E,
\end{align*}
where $\ket{\phi_0}$ and $\ket{\phi_1}$ are two different non-orthogonal states. Bob can guess Alice's key without an error by measuring his state in the computational basis. 
His ability to predict the conjugate basis will depend on the overlap of $\ket{\phi_0}$ and $\ket{\phi_1}$. Assuming this is not nearly maximal, Alice will have to provide Bob with some additional information, which in this case would be
the result of measuring some set of stabilizers. Measuring two stabilizers defeats their purpose, since then no
secret key can be distilled. Hence Alice measures either $X \otimes X$,  $X \otimes \mathbbm{1}$ or $\mathbbm{1} \otimes X$. The case where  
 $X \otimes \mathbbm{1}$ or $\mathbbm{1} \otimes X$ is used simply reduces to the case in which Alice and Bob only share one state $\ket{\psi}^{ABSE}$. In that case, Bob's minimum error probability of guessing Alice's measurement in the conjugated basis is given by $\frac{1}{2}-\frac{1}{2}\sqrt{1-|\bracket{\phi_0}{\phi_1}|^2}$ (which follows from Helstrom's result~\cite{helstrom_quantum_1976} for pure states) and the measurement used is independent of $\beta$. 
However, if $X \otimes X$ is used instead, then the minimum error probability of the optimal measurement given any $\beta$ is $\frac{1}{2}-\frac{1}{2}\sqrt{1-|\bracket{\phi_0}{\phi_1}|^4}$, which is smaller than $\frac{1}{2}-\frac{1}{2}\sqrt{1-|\bracket{\phi_0}{\phi_1}|^2}$. For each value of $\beta$, the optimal measurement is, for $\beta=0$, the two projections $\hat{P}^{\beta=0}_{\pm}$ on the range of the positive and negative parts of $(\phi^S_0)^{\otimes 2}-(\phi^S_1)^{\otimes 2}$ and the extra projection so that the sum of them is $\mathbbm{1}$. 
For $\beta=1$, the optimal measurement is the two projections $\hat{P}^{\beta=1}_{\pm}$ on the range of the positive and negative parts of $\phi^S_0 \otimes \phi^S_1-\phi^S_1\otimes \phi^S_0$ added with an extra projection so that the sum of them is $\mathbbm{1}$. Since the projection $\hat{P}^{\beta=0}_{+}$ overlaps with both $\hat{P}^{\beta=1}_{\pm}$, the optimal measurement $\widetilde{\Lambda}^{BS}_{\beta}$ cannot be independent of $\beta$.

Despite this example, we have not shown that the asymptotic rates of some protocols (for $n\rightarrow \infty$) could not be achieved using a measurement $\widetilde{\Lambda}^{BS}$ that is independent of $\beta$, but it seems reasonable to conjuncture that this is the case. Even if it were unnecessary, allowing $\widetilde{\Lambda}^{BS}$ to depend on $\beta$ does help to prove Theorems~\ref{thm:dp} and~\ref{th:HashIneq}.

\bibliography{psd}

\begin{thebibliography}{58}
\expandafter\ifx\csname natexlab\endcsname\relax\def\natexlab#1{#1}\fi
\expandafter\ifx\csname bibnamefont\endcsname\relax
  \def\bibnamefont#1{#1}\fi
\expandafter\ifx\csname bibfnamefont\endcsname\relax
  \def\bibfnamefont#1{#1}\fi
\expandafter\ifx\csname citenamefont\endcsname\relax
  \def\citenamefont#1{#1}\fi
\expandafter\ifx\csname url\endcsname\relax
  \def\url#1{\texttt{#1}}\fi
\expandafter\ifx\csname urlprefix\endcsname\relax\def\urlprefix{URL }\fi
\providecommand{\bibinfo}[2]{#2}
\providecommand{\eprint}[2][]{\url{#2}}

\bibitem[{\citenamefont{Mayers}(1996)}]{mayers_quantum_1996}
\bibinfo{author}{\bibfnamefont{D.}~\bibnamefont{Mayers}}, in
  \emph{\bibinfo{booktitle}{Advances in Cryptology --- CRYPTO '96}}
  (\bibinfo{publisher}{Springer}, \bibinfo{year}{1996}), vol.
  \bibinfo{volume}{1109/1996} of \emph{\bibinfo{series}{Lecture Notes in
  Computer Science}}, pp. \bibinfo{pages}{343--357}.

\bibitem[{\citenamefont{Lo and Chau}(1999)}]{lo_unconditional_1999}
\bibinfo{author}{\bibfnamefont{H.-K.} \bibnamefont{Lo}} \bibnamefont{and}
  \bibinfo{author}{\bibfnamefont{H.~F.} \bibnamefont{Chau}},
  \bibinfo{journal}{Science} \textbf{\bibinfo{volume}{283}},
  \bibinfo{pages}{2050} (\bibinfo{year}{1999}).

\bibitem[{\citenamefont{Shor and Preskill}(2000)}]{shor_simple_2000}
\bibinfo{author}{\bibfnamefont{P.~W.} \bibnamefont{Shor}} \bibnamefont{and}
  \bibinfo{author}{\bibfnamefont{J.}~\bibnamefont{Preskill}},
  \bibinfo{journal}{Phys. Rev. Lett.} \textbf{\bibinfo{volume}{85}},
  \bibinfo{pages}{441} (\bibinfo{year}{2000}).

\bibitem[{\citenamefont{Lo}(2001)}]{lo_proof_2001}
\bibinfo{author}{\bibfnamefont{H.-K.} \bibnamefont{Lo}},
  \bibinfo{journal}{Quantum Inf. Comput.} \textbf{\bibinfo{volume}{1}},
  \bibinfo{pages}{81} (\bibinfo{year}{2001}).

\bibitem[{\citenamefont{Tamaki et~al.}(2003)\citenamefont{Tamaki, Koashi, and
  Imoto}}]{tamaki_unconditionally_2003}
\bibinfo{author}{\bibfnamefont{K.}~\bibnamefont{Tamaki}},
  \bibinfo{author}{\bibfnamefont{M.}~\bibnamefont{Koashi}}, \bibnamefont{and}
  \bibinfo{author}{\bibfnamefont{N.}~\bibnamefont{Imoto}},
  \bibinfo{journal}{Phys. Rev. Lett.} \textbf{\bibinfo{volume}{90}},
  \bibinfo{pages}{167904} (\bibinfo{year}{2003}).

\bibitem[{\citenamefont{Boileau et~al.}(2005)\citenamefont{Boileau, Tamaki,
  Batuwantudawe, Laflamme, and Renes}}]{boileau_unconditional_2005}
\bibinfo{author}{\bibfnamefont{J.-C.} \bibnamefont{Boileau}},
  \bibinfo{author}{\bibfnamefont{K.}~\bibnamefont{Tamaki}},
  \bibinfo{author}{\bibfnamefont{J.}~\bibnamefont{Batuwantudawe}},
  \bibinfo{author}{\bibfnamefont{R.}~\bibnamefont{Laflamme}}, \bibnamefont{and}
  \bibinfo{author}{\bibfnamefont{J.~M.} \bibnamefont{Renes}},
  \bibinfo{journal}{Phys. Rev. Lett.} \textbf{\bibinfo{volume}{94}},
  \bibinfo{pages}{040503} (\bibinfo{year}{2005}).

\bibitem[{\citenamefont{Tamaki and Lo}(2006)}]{tamaki_unconditionally_2006}
\bibinfo{author}{\bibfnamefont{K.}~\bibnamefont{Tamaki}} \bibnamefont{and}
  \bibinfo{author}{\bibfnamefont{H.-K.} \bibnamefont{Lo}},
  \bibinfo{journal}{Phys. Rev. A} \textbf{\bibinfo{volume}{73}},
  \bibinfo{pages}{010302} (\bibinfo{year}{2006}).

\bibitem[{\citenamefont{Renes and Grassl}(2006)}]{renes_generalized_2006}
\bibinfo{author}{\bibfnamefont{J.~M.} \bibnamefont{Renes}} \bibnamefont{and}
  \bibinfo{author}{\bibfnamefont{M.}~\bibnamefont{Grassl}},
  \bibinfo{journal}{Phys. Rev. A} \textbf{\bibinfo{volume}{74}},
  \bibinfo{pages}{022317} (\bibinfo{year}{2006}).

\bibitem[{\citenamefont{Gottesman et~al.}(2004)\citenamefont{Gottesman, Lo,
  L{\"u}tkenhaus, and Preskill}}]{gottesman_security_2004}
\bibinfo{author}{\bibfnamefont{D.}~\bibnamefont{Gottesman}},
  \bibinfo{author}{\bibfnamefont{H.-K.} \bibnamefont{Lo}},
  \bibinfo{author}{\bibfnamefont{N.}~\bibnamefont{L{\"u}tkenhaus}},
  \bibnamefont{and} \bibinfo{author}{\bibfnamefont{J.}~\bibnamefont{Preskill}},
  \bibinfo{journal}{Quantum Inf. Comput.} \textbf{\bibinfo{volume}{4}},
  \bibinfo{pages}{325} (\bibinfo{year}{2004}).

\bibitem[{\citenamefont{Koashi}(2006{\natexlab{a}})}]{koashi_unconditional_200%
6}
\bibinfo{author}{\bibfnamefont{M.}~\bibnamefont{Koashi}}, \bibinfo{journal}{J.
  Phys.: Conf. Ser.} \textbf{\bibinfo{volume}{36}}, \bibinfo{pages}{98}
  (\bibinfo{year}{2006}{\natexlab{a}}).

\bibitem[{\citenamefont{Koashi}(2006{\natexlab{b}})}]{koashi_efficient_2006}
\bibinfo{author}{\bibfnamefont{M.}~\bibnamefont{Koashi}},
  \bibinfo{journal}{arXiv:quant-ph/0609180v1}
  (\bibinfo{year}{2006}{\natexlab{b}}).

\bibitem[{\citenamefont{Csiszar and Korner}(1978)}]{csiszar_broadcast_1978}
\bibinfo{author}{\bibfnamefont{I.}~\bibnamefont{Csiszar}} \bibnamefont{and}
  \bibinfo{author}{\bibfnamefont{J.}~\bibnamefont{Korner}},
  \bibinfo{journal}{IEEE Trans. Inf. Theory} \textbf{\bibinfo{volume}{24}},
  \bibinfo{pages}{339} (\bibinfo{year}{1978}).

\bibitem[{\citenamefont{Devetak and Winter}(2005)}]{devetak_distillation_2005}
\bibinfo{author}{\bibfnamefont{I.}~\bibnamefont{Devetak}} \bibnamefont{and}
  \bibinfo{author}{\bibfnamefont{A.}~\bibnamefont{Winter}},
  \bibinfo{journal}{Proc. R. Soc. A} \textbf{\bibinfo{volume}{461}},
  \bibinfo{pages}{207} (\bibinfo{year}{2005}).

\bibitem[{\citenamefont{Renner and K{\"o}nig}(2005)}]{renner_universally_2005}
\bibinfo{author}{\bibfnamefont{R.}~\bibnamefont{Renner}} \bibnamefont{and}
  \bibinfo{author}{\bibfnamefont{R.}~\bibnamefont{K{\"o}nig}}, in
  \emph{\bibinfo{booktitle}{Second Theory of Cryptography Conference}}
  (\bibinfo{publisher}{Springer}, \bibinfo{address}{Cambridge, MA},
  \bibinfo{year}{2005}), vol. \bibinfo{volume}{3378}, pp.
  \bibinfo{pages}{407--425}.

\bibitem[{\citenamefont{Kraus et~al.}(2005)\citenamefont{Kraus, Gisin, and
  Renner}}]{kraus_lower_2005}
\bibinfo{author}{\bibfnamefont{B.}~\bibnamefont{Kraus}},
  \bibinfo{author}{\bibfnamefont{N.}~\bibnamefont{Gisin}}, \bibnamefont{and}
  \bibinfo{author}{\bibfnamefont{R.}~\bibnamefont{Renner}},
  \bibinfo{journal}{Phys. Rev. Lett.} \textbf{\bibinfo{volume}{95}},
  \bibinfo{pages}{080501} (\bibinfo{year}{2005}).

\bibitem[{\citenamefont{Renner et~al.}(2005)\citenamefont{Renner, Gisin, and
  Kraus}}]{renner_information-theoretic_2005}
\bibinfo{author}{\bibfnamefont{R.}~\bibnamefont{Renner}},
  \bibinfo{author}{\bibfnamefont{N.}~\bibnamefont{Gisin}}, \bibnamefont{and}
  \bibinfo{author}{\bibfnamefont{B.}~\bibnamefont{Kraus}},
  \bibinfo{journal}{Phys. Rev. A} \textbf{\bibinfo{volume}{72}},
  \bibinfo{pages}{012332} (\bibinfo{year}{2005}).

\bibitem[{\citenamefont{Renner}(2006)}]{renner_security_2006}
\bibinfo{author}{\bibfnamefont{R.}~\bibnamefont{Renner}}, Ph.D. thesis,
  \bibinfo{school}{ETH Z{\"u}rich} (\bibinfo{year}{2006}).

\bibitem[{\citenamefont{Horodecki et~al.}(2005)\citenamefont{Horodecki,
  Horodecki, Horodecki, and Oppenheim}}]{horodecki_secure_2005}
\bibinfo{author}{\bibfnamefont{K.}~\bibnamefont{Horodecki}},
  \bibinfo{author}{\bibfnamefont{M.}~\bibnamefont{Horodecki}},
  \bibinfo{author}{\bibfnamefont{P.}~\bibnamefont{Horodecki}},
  \bibnamefont{and}
  \bibinfo{author}{\bibfnamefont{J.}~\bibnamefont{Oppenheim}},
  \bibinfo{journal}{Phys. Rev. Lett.} \textbf{\bibinfo{volume}{94}},
  \bibinfo{pages}{160502} (\bibinfo{year}{2005}).

\bibitem[{\citenamefont{Horodecki
  et~al.}(2008{\natexlab{a}})\citenamefont{Horodecki, Horodecki, Horodecki, and
  Oppenheim}}]{horodecki_general_2005}
\bibinfo{author}{\bibfnamefont{K.}~\bibnamefont{Horodecki}},
  \bibinfo{author}{\bibfnamefont{M.}~\bibnamefont{Horodecki}},
  \bibinfo{author}{\bibfnamefont{P.}~\bibnamefont{Horodecki}},
  \bibnamefont{and}
  \bibinfo{author}{\bibfnamefont{J.}~\bibnamefont{Oppenheim}},
  \bibinfo{journal}{arXiv:quant-ph/0506189v2}
  (\bibinfo{year}{2008}{\natexlab{a}}).

\bibitem[{\citenamefont{Calderbank and Shor}(1996)}]{calderbank_good_1996}
\bibinfo{author}{\bibfnamefont{A.~R.} \bibnamefont{Calderbank}}
  \bibnamefont{and} \bibinfo{author}{\bibfnamefont{P.~W.} \bibnamefont{Shor}},
  \bibinfo{journal}{Phys. Rev. A} \textbf{\bibinfo{volume}{54}},
  \bibinfo{pages}{1098} (\bibinfo{year}{1996}).

\bibitem[{\citenamefont{Steane}(1996)}]{steane_multiple-particle_1996}
\bibinfo{author}{\bibfnamefont{A.}~\bibnamefont{Steane}},
  \bibinfo{journal}{Proc. R. Soc. A} \textbf{\bibinfo{volume}{452}},
  \bibinfo{pages}{2551} (\bibinfo{year}{1996}).

\bibitem[{\citenamefont{Koashi}(2007)}]{koashi_complementarity_2007}
\bibinfo{author}{\bibfnamefont{M.}~\bibnamefont{Koashi}},
  \bibinfo{journal}{arXiv:0704.3661v1 [quant-ph]}  (\bibinfo{year}{2007}).

\bibitem[{\citenamefont{Holevo}(1998)}]{holevo_capacity_1998}
\bibinfo{author}{\bibfnamefont{A.}~\bibnamefont{Holevo}},
  \bibinfo{journal}{IEEE Trans. Inf. Theory} \textbf{\bibinfo{volume}{44}},
  \bibinfo{pages}{269} (\bibinfo{year}{1998}).

\bibitem[{\citenamefont{Schumacher and
  Westmoreland}(1997)}]{schumacher_sending_1997}
\bibinfo{author}{\bibfnamefont{B.}~\bibnamefont{Schumacher}} \bibnamefont{and}
  \bibinfo{author}{\bibfnamefont{M.~D.} \bibnamefont{Westmoreland}},
  \bibinfo{journal}{Phys. Rev. A} \textbf{\bibinfo{volume}{56}},
  \bibinfo{pages}{131} (\bibinfo{year}{1997}).

\bibitem[{\citenamefont{Deutsch et~al.}(1996)\citenamefont{Deutsch, Ekert,
  Jozsa, Macchiavello, Popescu, and Sanpera}}]{deutsch_quantum_1996}
\bibinfo{author}{\bibfnamefont{D.}~\bibnamefont{Deutsch}},
  \bibinfo{author}{\bibfnamefont{A.}~\bibnamefont{Ekert}},
  \bibinfo{author}{\bibfnamefont{R.}~\bibnamefont{Jozsa}},
  \bibinfo{author}{\bibfnamefont{C.}~\bibnamefont{Macchiavello}},
  \bibinfo{author}{\bibfnamefont{S.}~\bibnamefont{Popescu}}, \bibnamefont{and}
  \bibinfo{author}{\bibfnamefont{A.}~\bibnamefont{Sanpera}},
  \bibinfo{journal}{Phys. Rev. Lett.} \textbf{\bibinfo{volume}{77}},
  \bibinfo{pages}{2818} (\bibinfo{year}{1996}).

\bibitem[{\citenamefont{Schumacher and
  Westmoreland}(1998)}]{schumacher_quantum_1998}
\bibinfo{author}{\bibfnamefont{B.}~\bibnamefont{Schumacher}} \bibnamefont{and}
  \bibinfo{author}{\bibfnamefont{M.~D.} \bibnamefont{Westmoreland}},
  \bibinfo{journal}{Phys. Rev. Lett.} \textbf{\bibinfo{volume}{80}},
  \bibinfo{pages}{5695} (\bibinfo{year}{1998}).

\bibitem[{\citenamefont{Devetak and Winter}(2004)}]{devetak_relating_2004}
\bibinfo{author}{\bibfnamefont{I.}~\bibnamefont{Devetak}} \bibnamefont{and}
  \bibinfo{author}{\bibfnamefont{A.}~\bibnamefont{Winter}},
  \bibinfo{journal}{Phys. Rev. Lett.} \textbf{\bibinfo{volume}{93}},
  \bibinfo{pages}{080501} (\bibinfo{year}{2004}).

\bibitem[{\citenamefont{Maassen and Uffink}(1988)}]{maassen_generalized_1988}
\bibinfo{author}{\bibfnamefont{H.}~\bibnamefont{Maassen}} \bibnamefont{and}
  \bibinfo{author}{\bibfnamefont{J.~B.~M.} \bibnamefont{Uffink}},
  \bibinfo{journal}{Phys. Rev. Lett.} \textbf{\bibinfo{volume}{60}},
  \bibinfo{pages}{1103} (\bibinfo{year}{1988}).

\bibitem[{\citenamefont{Hall}(1995)}]{hall_information_1995}
\bibinfo{author}{\bibfnamefont{M.~J.~W.} \bibnamefont{Hall}},
  \bibinfo{journal}{Phys. Rev. Lett.} \textbf{\bibinfo{volume}{74}},
  \bibinfo{pages}{3307} (\bibinfo{year}{1995}).

\bibitem[{\citenamefont{Uhlmann}(1976)}]{uhlmann_transition_1976}
\bibinfo{author}{\bibfnamefont{A.}~\bibnamefont{Uhlmann}},
  \bibinfo{journal}{Rep. Math. Phys.} \textbf{\bibinfo{volume}{9}},
  \bibinfo{pages}{273} (\bibinfo{year}{1976}).

\bibitem[{\citenamefont{Jozsa}(1994)}]{jozsa_fidelity_1994}
\bibinfo{author}{\bibfnamefont{R.}~\bibnamefont{Jozsa}}, \bibinfo{journal}{J.
  Mod. Opt.} \textbf{\bibinfo{volume}{41}}, \bibinfo{pages}{2315}
  (\bibinfo{year}{1994}).

\bibitem[{\citenamefont{Cerf et~al.}(2002)\citenamefont{Cerf, Bourennane,
  Karlsson, and Gisin}}]{cerf_security_2002}
\bibinfo{author}{\bibfnamefont{N.~J.} \bibnamefont{Cerf}},
  \bibinfo{author}{\bibfnamefont{M.}~\bibnamefont{Bourennane}},
  \bibinfo{author}{\bibfnamefont{A.}~\bibnamefont{Karlsson}}, \bibnamefont{and}
  \bibinfo{author}{\bibfnamefont{N.}~\bibnamefont{Gisin}},
  \bibinfo{journal}{Phys. Rev. Lett.} \textbf{\bibinfo{volume}{88}},
  \bibinfo{pages}{127902} (\bibinfo{year}{2002}).

\bibitem[{\citenamefont{Helstrom}(1976)}]{helstrom_quantum_1976}
\bibinfo{author}{\bibfnamefont{C.~W.} \bibnamefont{Helstrom}},
  \emph{\bibinfo{title}{Quantum detection and estimation theory}}, vol.
  \bibinfo{volume}{123} of \emph{\bibinfo{series}{Mathematics in Science and
  Engineering}} (\bibinfo{publisher}{Academic}, \bibinfo{address}{London},
  \bibinfo{year}{1976}).

\bibitem[{\citenamefont{Ben-Or et~al.}(2005)\citenamefont{Ben-Or, Horodecki,
  Leung, Mayers, and Oppenheim}}]{ben-or_universal_2005}
\bibinfo{author}{\bibfnamefont{M.}~\bibnamefont{Ben-Or}},
  \bibinfo{author}{\bibfnamefont{M.}~\bibnamefont{Horodecki}},
  \bibinfo{author}{\bibfnamefont{D.~W.} \bibnamefont{Leung}},
  \bibinfo{author}{\bibfnamefont{D.}~\bibnamefont{Mayers}}, \bibnamefont{and}
  \bibinfo{author}{\bibfnamefont{J.}~\bibnamefont{Oppenheim}}, in
  \emph{\bibinfo{booktitle}{Second Theory of Cryptography Conference}}
  (\bibinfo{publisher}{Springer}, \bibinfo{address}{Cambridge, MA},
  \bibinfo{year}{2005}), vol. \bibinfo{volume}{3378}, pp.
  \bibinfo{pages}{386--406}.

\bibitem[{\citenamefont{Gottesman and Lo}(2003)}]{gottesman_proof_2003}
\bibinfo{author}{\bibfnamefont{D.}~\bibnamefont{Gottesman}} \bibnamefont{and}
  \bibinfo{author}{\bibfnamefont{H.-K.} \bibnamefont{Lo}},
  \bibinfo{journal}{IEEE Trans. Inf. Theory} \textbf{\bibinfo{volume}{49}},
  \bibinfo{pages}{457} (\bibinfo{year}{2003}).

\bibitem[{\citenamefont{Gottesman et~al.}(2001)\citenamefont{Gottesman, Kitaev,
  and Preskill}}]{gottesman_encoding_2001}
\bibinfo{author}{\bibfnamefont{D.}~\bibnamefont{Gottesman}},
  \bibinfo{author}{\bibfnamefont{A.}~\bibnamefont{Kitaev}}, \bibnamefont{and}
  \bibinfo{author}{\bibfnamefont{J.}~\bibnamefont{Preskill}},
  \bibinfo{journal}{Phys. Rev. A} \textbf{\bibinfo{volume}{64}},
  \bibinfo{pages}{012310} (\bibinfo{year}{2001}).

\bibitem[{\citenamefont{Fuchs and van~de
  Graaf}(1999)}]{fuchs_cryptographic_1999}
\bibinfo{author}{\bibfnamefont{C.}~\bibnamefont{Fuchs}} \bibnamefont{and}
  \bibinfo{author}{\bibfnamefont{J.}~\bibnamefont{van~de Graaf}},
  \bibinfo{journal}{IEEE Trans. Inf. Theory} \textbf{\bibinfo{volume}{45}},
  \bibinfo{pages}{1216} (\bibinfo{year}{1999}).

\bibitem[{\citenamefont{Renes and Smith}(2007)}]{renes_noisy_2007}
\bibinfo{author}{\bibfnamefont{J.~M.} \bibnamefont{Renes}} \bibnamefont{and}
  \bibinfo{author}{\bibfnamefont{G.}~\bibnamefont{Smith}},
  \bibinfo{journal}{Phys. Rev. Lett.} \textbf{\bibinfo{volume}{98}},
  \bibinfo{pages}{020502} (\bibinfo{year}{2007}).

\bibitem[{\citenamefont{DiVincenzo et~al.}(2004)\citenamefont{DiVincenzo,
  Horodecki, Leung, Smolin, and Terhal}}]{divincenzo_locking_2004}
\bibinfo{author}{\bibfnamefont{D.~P.} \bibnamefont{DiVincenzo}},
  \bibinfo{author}{\bibfnamefont{M.}~\bibnamefont{Horodecki}},
  \bibinfo{author}{\bibfnamefont{D.~W.} \bibnamefont{Leung}},
  \bibinfo{author}{\bibfnamefont{J.~A.} \bibnamefont{Smolin}},
  \bibnamefont{and} \bibinfo{author}{\bibfnamefont{B.~M.}
  \bibnamefont{Terhal}}, \bibinfo{journal}{Phys. Rev. Lett.}
  \textbf{\bibinfo{volume}{92}}, \bibinfo{pages}{067902}
  (\bibinfo{year}{2004}).

\bibitem[{\citenamefont{Lo}(2003)}]{lo_method_2003}
\bibinfo{author}{\bibfnamefont{H.-K.} \bibnamefont{Lo}}, \bibinfo{journal}{New
  J. Phys.} \textbf{\bibinfo{volume}{5}}, \bibinfo{pages}{36}
  (\bibinfo{year}{2003}).

\bibitem[{\citenamefont{Horodecki
  et~al.}(2008{\natexlab{b}})\citenamefont{Horodecki, Leung, and
  Oppenheim}}]{horodecki_quantum_2006}
\bibinfo{author}{\bibfnamefont{K.}~\bibnamefont{Horodecki}},
  \bibinfo{author}{\bibfnamefont{D.}~\bibnamefont{Leung}}, \bibnamefont{and}
  \bibinfo{author}{\bibfnamefont{J.}~\bibnamefont{Oppenheim}},
  \bibinfo{journal}{IEEE Trans. Inf. Theory} \textbf{\bibinfo{volume}{54}},
  \bibinfo{pages}{2604} (\bibinfo{year}{2008}{\natexlab{b}}).

\bibitem[{\citenamefont{Lloyd}(1997)}]{lloyd_capacity_1997}
\bibinfo{author}{\bibfnamefont{S.}~\bibnamefont{Lloyd}},
  \bibinfo{journal}{Phys. Rev. A} \textbf{\bibinfo{volume}{55}},
  \bibinfo{pages}{1613} (\bibinfo{year}{1997}).

\bibitem[{\citenamefont{Shor}(2002)}]{shor_quantum_2002}
\bibinfo{author}{\bibfnamefont{P.~W.} \bibnamefont{Shor}},
  \emph{\bibinfo{title}{The quantum channel capacity and coherent
  information}}, \bibinfo{howpublished}{MSRI Seminar} (\bibinfo{year}{2002}).

\bibitem[{\citenamefont{Devetak}(2005)}]{devetak_private_2005}
\bibinfo{author}{\bibfnamefont{I.}~\bibnamefont{Devetak}},
  \bibinfo{journal}{IEEE Trans. Inf. Theory} \textbf{\bibinfo{volume}{51}},
  \bibinfo{pages}{44} (\bibinfo{year}{2005}).

\bibitem[{\citenamefont{Horodecki
  et~al.}(2008{\natexlab{c}})\citenamefont{Horodecki, Lloyd, and
  Winter}}]{horodecki_quantum_2007}
\bibinfo{author}{\bibfnamefont{M.}~\bibnamefont{Horodecki}},
  \bibinfo{author}{\bibfnamefont{S.}~\bibnamefont{Lloyd}}, \bibnamefont{and}
  \bibinfo{author}{\bibfnamefont{A.}~\bibnamefont{Winter}},
  \bibinfo{journal}{Open Syst. Inf. Dyn.} \textbf{\bibinfo{volume}{15}},
  \bibinfo{pages}{47} (\bibinfo{year}{2008}{\natexlab{c}}).

\bibitem[{\citenamefont{Hayden et~al.}(2008{\natexlab{a}})\citenamefont{Hayden,
  Horodecki, Yard, and Winter}}]{hayden_decoupling_2007}
\bibinfo{author}{\bibfnamefont{P.}~\bibnamefont{Hayden}},
  \bibinfo{author}{\bibfnamefont{M.}~\bibnamefont{Horodecki}},
  \bibinfo{author}{\bibfnamefont{J.}~\bibnamefont{Yard}}, \bibnamefont{and}
  \bibinfo{author}{\bibfnamefont{A.}~\bibnamefont{Winter}},
  \bibinfo{journal}{Open Syst. Inf. Dyn.} \textbf{\bibinfo{volume}{15}},
  \bibinfo{pages}{7} (\bibinfo{year}{2008}{\natexlab{a}}).

\bibitem[{\citenamefont{Hayden et~al.}(2008{\natexlab{b}})\citenamefont{Hayden,
  Shor, and Winter}}]{hayden_random_2007}
\bibinfo{author}{\bibfnamefont{P.}~\bibnamefont{Hayden}},
  \bibinfo{author}{\bibfnamefont{P.~W.} \bibnamefont{Shor}}, \bibnamefont{and}
  \bibinfo{author}{\bibfnamefont{A.}~\bibnamefont{Winter}},
  \bibinfo{journal}{Open Syst. Inf. Dyn.} \textbf{\bibinfo{volume}{15}},
  \bibinfo{pages}{71} (\bibinfo{year}{2008}{\natexlab{b}}).

\bibitem[{\citenamefont{Kretschmann et~al.}(2008)\citenamefont{Kretschmann,
  Schlingemann, and Werner}}]{kretschmann_information-disturbance_2006}
\bibinfo{author}{\bibfnamefont{D.}~\bibnamefont{Kretschmann}},
  \bibinfo{author}{\bibfnamefont{D.}~\bibnamefont{Schlingemann}},
  \bibnamefont{and} \bibinfo{author}{\bibfnamefont{R.~F.}
  \bibnamefont{Werner}}, \bibinfo{journal}{IEEE Trans. Inf. Theory}
  \textbf{\bibinfo{volume}{54}}, \bibinfo{pages}{1708} (\bibinfo{year}{2008}).

\bibitem[{\citenamefont{Kretschmann
  et~al.}(2007{\natexlab{a}})\citenamefont{Kretschmann, Schlingemann, and
  Werner}}]{kretschmann_continuity_2007}
\bibinfo{author}{\bibfnamefont{D.}~\bibnamefont{Kretschmann}},
  \bibinfo{author}{\bibfnamefont{D.}~\bibnamefont{Schlingemann}},
  \bibnamefont{and} \bibinfo{author}{\bibfnamefont{R.~F.}
  \bibnamefont{Werner}}, \bibinfo{journal}{arXiv:0710.2495v1 [quant-ph]}
  (\bibinfo{year}{2007}{\natexlab{a}}).

\bibitem[{\citenamefont{Kretschmann
  et~al.}(2007{\natexlab{b}})\citenamefont{Kretschmann, Kribs, and
  Spekkens}}]{kretschmann_complementarity_2007}
\bibinfo{author}{\bibfnamefont{D.}~\bibnamefont{Kretschmann}},
  \bibinfo{author}{\bibfnamefont{D.~W.} \bibnamefont{Kribs}}, \bibnamefont{and}
  \bibinfo{author}{\bibfnamefont{R.~W.} \bibnamefont{Spekkens}},
  \bibinfo{journal}{arXiv:0711.3438v1 [quant-ph]}
  (\bibinfo{year}{2007}{\natexlab{b}}).

\bibitem[{\citenamefont{Inoue et~al.}(2002)\citenamefont{Inoue, Waks, and
  Yamamoto}}]{inoue_differential_2002}
\bibinfo{author}{\bibfnamefont{K.}~\bibnamefont{Inoue}},
  \bibinfo{author}{\bibfnamefont{E.}~\bibnamefont{Waks}}, \bibnamefont{and}
  \bibinfo{author}{\bibfnamefont{Y.}~\bibnamefont{Yamamoto}},
  \bibinfo{journal}{Phys. Rev. Lett.} \textbf{\bibinfo{volume}{89}},
  \bibinfo{pages}{037902} (\bibinfo{year}{2002}).

\bibitem[{\citenamefont{Stucki et~al.}(2005)\citenamefont{Stucki, Brunner,
  Gisin, Scarani, and Zbinden}}]{stucki_fast_2005}
\bibinfo{author}{\bibfnamefont{D.}~\bibnamefont{Stucki}},
  \bibinfo{author}{\bibfnamefont{N.}~\bibnamefont{Brunner}},
  \bibinfo{author}{\bibfnamefont{N.}~\bibnamefont{Gisin}},
  \bibinfo{author}{\bibfnamefont{V.}~\bibnamefont{Scarani}}, \bibnamefont{and}
  \bibinfo{author}{\bibfnamefont{H.}~\bibnamefont{Zbinden}},
  \bibinfo{journal}{Appl. Phys. Lett.} \textbf{\bibinfo{volume}{87}},
  \bibinfo{pages}{194108} (\bibinfo{year}{2005}).

\bibitem[{\citenamefont{Renes and Boileau}(2008)}]{renes_strong_2008}
\bibinfo{author}{\bibfnamefont{J.~M.} \bibnamefont{Renes}} \bibnamefont{and}
  \bibinfo{author}{\bibfnamefont{J.-C.} \bibnamefont{Boileau}},
  \bibinfo{journal}{arXiv:0806.3984 [quant-ph]}  (\bibinfo{year}{2008}).

\bibitem[{\citenamefont{Winter}(1999)}]{winter_coding_1999}
\bibinfo{author}{\bibfnamefont{A.}~\bibnamefont{Winter}}, Ph.D. thesis,
  \bibinfo{school}{Universit{\"a}t Bielefeld} (\bibinfo{year}{1999}).

\bibitem[{\citenamefont{Devetak and Winter}(2003)}]{devetak_classical_2003}
\bibinfo{author}{\bibfnamefont{I.}~\bibnamefont{Devetak}} \bibnamefont{and}
  \bibinfo{author}{\bibfnamefont{A.}~\bibnamefont{Winter}},
  \bibinfo{journal}{Phys. Rev. A} \textbf{\bibinfo{volume}{68}},
  \bibinfo{pages}{042301} (\bibinfo{year}{2003}).

\bibitem[{\citenamefont{Hausladen and Wootters}(1994)}]{hausladen_pretty_1994}
\bibinfo{author}{\bibfnamefont{P.}~\bibnamefont{Hausladen}} \bibnamefont{and}
  \bibinfo{author}{\bibfnamefont{W.~K.} \bibnamefont{Wootters}},
  \bibinfo{journal}{J. Mod. Opt.} \textbf{\bibinfo{volume}{41}},
  \bibinfo{pages}{2385} (\bibinfo{year}{1994}).

\bibitem[{\citenamefont{Carter and Wegman}(1979)}]{carter_universal_1979}
\bibinfo{author}{\bibfnamefont{J.~L.} \bibnamefont{Carter}} \bibnamefont{and}
  \bibinfo{author}{\bibfnamefont{M.~N.} \bibnamefont{Wegman}},
  \bibinfo{journal}{J. Comput. Syst. Sci.} \textbf{\bibinfo{volume}{18}},
  \bibinfo{pages}{143} (\bibinfo{year}{1979}).

\bibitem[{\citenamefont{Cover and Thomas}(1991)}]{cover_elements_1991}
\bibinfo{author}{\bibfnamefont{T.~M.} \bibnamefont{Cover}} \bibnamefont{and}
  \bibinfo{author}{\bibfnamefont{J.~A.} \bibnamefont{Thomas}},
  \emph{\bibinfo{title}{Elements of Information Theory}}
  (\bibinfo{publisher}{Wiley-Interscience}, \bibinfo{year}{1991}).

\end{thebibliography}

\end{document}